\documentclass[10pt, journal]{IEEEtran}
\IEEEoverridecommandlockouts
\usepackage[noadjust]{cite}
\usepackage{amsmath, amssymb, amsfonts, url}
\usepackage[ruled,vlined]{algorithm2e}
\usepackage{graphicx, tabularx, booktabs, subcaption}
\usepackage{textcomp}
\usepackage{xcolor}
\usepackage[acronym]{glossaries}

\DeclareMathOperator*{\argmin}{arg\min}
\def\BibTeX{{\rm B\kern-.05em{\sc i\kern-.025em b}\kern-.08em
    T\kern-.1667em\lower.7ex\hbox{E}\kern-.125emX}}

\begin{document}

\title{Before AI Takes Over: Rethinking Nonlinear Signal Processing in Communications
\thanks{This work is part of the project SOFIA PID2023-147305OB-C32 funded by MICIU/AEI/10.13039/501100011033 and FEDER/UE.}
}

\author{\IEEEauthorblockN{Ana Pérez-Neira\IEEEauthorrefmark{1}\IEEEauthorrefmark{2}\IEEEauthorrefmark{3}, Marc Martinez-Gost\IEEEauthorrefmark{1}\IEEEauthorrefmark{2}, 
Miguel Ángel Lagunas\IEEEauthorrefmark{2}\\}
\IEEEauthorblockA{
\IEEEauthorrefmark{1}Centre Tecnològic de Telecomunicacions de Catalunya, Spain\\
\IEEEauthorrefmark{2}Dept. of Signal Theory and Communications, Universitat Politècnica de Catalunya, Spain\\
\IEEEauthorrefmark{3}ICREA Acadèmia, Spain\\
\{aperez, mmartinez, m.a.lagunas\}@cttc.es
}}

\newacronym{AWGN}{AWGN}{additive white Gaussian noise}
\newacronym{AI}{AI}{Artificial Intelligence}
\newacronym{AR}{AR}{autoregressive}
\newacronym{ARMA}{ARMA}{autoregressive moving-average}
\newacronym{BER}{BER}{bit error rate}
\newacronym{DCT}{DCT}{discrete cosine transform}
\newacronym{DFE}{DFE}{decision feedback equalizer}
\newacronym{DFT}{DFT}{discrete Fourier transform}
\newacronym{DIR}{DIR}{desired impulse response}
\newacronym{ENN}{ENN}{expressive neural network}
\newacronym{FIR}{FIR}{finite impulse response}
\newacronym{FPGAs}{FPGAs}{field-programmable gate arrays}
\newacronym{GPUs}{GPUs}{graphics processing units}
\newacronym{IIR}{IIR}{infinite impulse response}
\newacronym{ISI}{ISI}{intersymbol interference}
\newacronym{LMS}{LMS}{least mean squares}
\newacronym{LTE}{LTE}{long term evolution}
\newacronym{MDIR}{MDIR}{Matched DIR}
\newacronym{MIMO}{MIMO}{multiple-input multiple-output}
\newacronym{ML}{ML}{maximum likelihood}
\newacronym{MSE}{MSE}{mean squared error}
\newacronym{MLP}{MLP}{multilayer perceptron}
\newacronym{OFDM}{OFDM}{orthogonal frequency division multiplexing}
\newacronym{PAM}{PAM}{pulse amplitude modulation}
\newacronym{PAPR}{PAPR}{peak-to-average power ratio} 
\newacronym{RF}{RF}{radio-frequency}
\newacronym{SNR}{SNR}{signal-to-noise ratio}
\newacronym{VoIP}{VoIP}{voice over Internet protocol}
\newacronym{VoLTE}{VoLTE}{voice over LTE}

\maketitle
 \begin{abstract}
 There is an urgent reflection on traditional nonlinear signal processing methods in communications before Artificial Intelligence (AI)  dominates the field. It implies a need to reassess or reinterpret established theories and tools, highlighting the tension between data-driven and model-based approaches. This paper calls for preserving valuable insights from classical signal processing while exploring how they can coexist or integrate with emerging AI methods.
 \end{abstract}

Over the past few years, wireless communications have been undergoing a profound transformation. Traditional model-driven designs, which for decades formed the backbone of communication standards, are increasingly being complemented (and in some cases replaced) by \gls{AI} \cite{6g_ai_native,dl_phy, nature_6g}. This trend stems from the remarkable success of AI in fields such as computer vision and natural language processing, where data-driven models have dramatically outperformed conventional methods. Communications, however, pose a different challenge. Unlike image or speech, the physical layer of communication systems  have long been characterized by precise mathematical models, well-established physical principles and rigorously defined performance metrics, whose optimization has led to the development of a solid corpus of theory for communication systems.
In consequence, while AI's value in optimizing communication resources is well established \cite{rrm_ai}, its application to the design of the physical layer remains more controversial.

The rise of task-oriented and semantic communications \cite{semantics} is now encouraging the development of so-called AI-native transceivers, where the communication channel is modeled as an additional layer within a neural network architecture that jointly represents both the transmitter and the receiver. Other approaches leverage neural networks to implement specific nonlinear processing tasks within traditional transceivers \cite{machine_learning_wireless}.
We argue that there is a significant gap between the current theoretical foundations and the recently proposed AI-based solutions. We contend that the field urgently needs a robust framework for nonlinear signal processing to help us better understand and manage the nonlinear behaviors in complex communication systems that current AI methods attempt to address purely through data. As AI continues to play an increasing role in tackling problems beyond analytical tractability, advancing nonlinear signal processing becomes essential to bridge and harmonize data-driven and theory-driven paradigms.

This paper introduces a formal nonlinear signal processing framework that bridges classical adaptive system theory and modern AI, built upon the \gls{DCT} as a functional approximation tool. Leveraging this foundation, the authors reconsider the design of the artificial neuron itself and propose an alternative model, where the neuron activation function adapts based on learning and follows a DCT model. In \cite{ENN}, this approach leads to the development of the Expressive Neural Network (ENN), a compact or shallow multilayer perceptron which can approximate multivariate functions with lower number of parameters than existing neural networks. Going further in the neuron model, in \cite{source_separation}, the conventional perceptron is replaced by a single adaptive nonlinear function of its input, modeled again through the DCT. This simple yet expressive parametrization enables the model to blindly separate nonlinearly coupled signals. In essence, the trainable model functions as a new type of adaptive neuron, deeply rooted in a signal processing tool: the DCT. Throughout this paper, we use the term DCT-based neuron interchangeably to refer to a DCT model trained to learn a specific nonlinearity. 

The goal of this work is to address the fundamental problem of estimating a nonlinear communication channel (either its direct or inverse response) by modeling its nonlinear behavior through a DCT-based neuron. The proposed approach provides an effective, low-complexity and interpretable framework for characterizing nonlinear channels, enabling accurate parameterization and analysis prior to subsequent signal processing techniques. Rather than offering an exhaustive comparison with alternative methods, this paper aims to present a perspective on how nonlinear transmission systems can be approached using classical signal processing principles. The proposed DCT-based modeling tool, which has already demonstrated promising results in complex scenarios, bridges the gap between conventional communication transceiver design and modern nonlinear environments.

The rest of the paper is organized as follows: Section I introduces and motivates the problem of nonlinear channels across various communication scenarios; Section II discusses the appropriate modeling of nonlinear functions and proposes the DCT-based estimation framework; Section III presents the proposed framework for nonlinear channel estimation in flat-fading scenarios, addressing both the inverse and direct estimation approaches; Section IV extends the framework to channels with memory and Section V concludes the paper.




\section{Towards a Unified Framework for Nonlinear Channel Models}

In every modern communication system, signals pass through a chain of physical components and channels that distort them in various ways. Some of these distortions are well understood and can be neatly described with linear models: a signal may be delayed, filtered or attenuated, and we can capture all of this mathematically with convolutions and impulse responses. Linear models are elegant and powerful because they preserve the principle of superposition, this is, a linear system does not introduce new frequencies that were not present at the input.

However, in practical communication systems, the equivalent channel that a signal experiences is rarely perfectly linear. Nonlinearities arise from both the hardware and the physical medium, and specialized nonlinear signal processing techniques are often required to mitigate these effects. Unlike linear systems, there is still no unified theory for nonlinear system behavior, and modeling approaches often depend on the dominant type of nonlinearity in the channel.

For instance, in wireless and \gls{RF} communication systems, power amplifiers operating near saturation introduce nonlinear multiplicative effects, where the amplitude or phase of the transmitted signal is distorted in proportion to its instantaneous power. Such channels can be modeled as nonlinear multiplicative channels, and techniques like digital predistortion are employed to compensate for these amplitude-dependent distortions \cite{PA_modeling}.

In \gls{MIMO} or time-dispersive channels, where the present output depends nonlinearly on previous outputs, a nonlinear and  \gls{AR} model provides a more accurate description. These models are useful for nonlinear detection and equalization in systems affected by \gls{ISI} or coupling between spatial streams \cite{mimo_survey}. Similarly, in \gls{OFDM} systems such as \gls{LTE} or Wi-Fi, where clipping distortion and high \gls{PAPR} create memory effects, nonlinear AR-based modeling helps describe the nonlinear evolution of the channel response over time \cite{papr_ofdm}.

Extending this idea further, nonlinear and \gls{ARMA} models capture both the nonlinear memory in past outputs and the influence of past inputs. This formulation is particularly relevant in satellite communications, where traveling-wave tube amplifiers (TWTAs) generate strong nonlinearities and memory effects due to multiple carriers and thermal dynamics. The resulting nonlinear intermodulation products can be effectively represented using nonlinear ARMA-like channel models, which support both transmitter-side predistortion \cite{satcom_pre} and receiver-side nonlinear equalization \cite{satcom_post}.

Similar nonlinear and memory-based effects arise in optical fiber systems, where Kerr nonlinearity induces self-phase modulation, cross-phase modulation and four-wave mixing. These complex behaviors can also be captured within the nonlinear ARMA framework, combining nonlinear dependencies with dispersive memory effects accumulated over long distances \cite{kerr_effect2}.

The drive for higher spectral efficiency in next-generation wireless networks has also promoted the development of full-duplex radios, which transmit and receive simultaneously on the same frequency band. Here, the equivalent self-interference channel exhibits both frequency selectivity and nonlinear multiplicative distortion due to reflections and amplifier nonlinearity. Advanced nonlinear self-interference cancellation methods are therefore required \cite{full_duplex1, full_duplex_2}.

Nonlinear mitigation is equally relevant in audio and speech transmission, where nonlinear echoes, noise, and reverberation distort the communication path between speaker and listener. Nonlinear AR models and nonlinear spatial filters are often used in modern microphone arrays to improve intelligibility \cite{speech_arrays, speech_arrays2}. Beyond conventional wireless and optical links, future networks must also support acoustic communication in complex environments such as underwater or ultrasonic bands, where multipath-induced frequency selectivity combines with nonlinear transducer and amplifier effects. These scenarios often require nonlinear  ARMA-based echo cancellation and nonlinear channel estimation to maintain reliable data transfer where linear methods fail \cite{underwater_comms, echo_w_linear, echo_volterra}.


Across all these scenarios, it becomes evident that an enormous variety of techniques has been developed to address nonlinear signal processing, each tailored to a specific domain and impairment. This diversity reveals a fragmented landscape that lacks a unified theoretical framework for nonlinear modeling and mitigation. In recent years, the rise of AI has been seen as a way to bridge these gaps, offering a common data-driven approach capable of tackling nearly all such problems, ranging from \gls{PAPR} reduction \cite{papr_ai} and optical fiber compensation \cite{kerr_ai} to speech enhancement \cite{speech_ai}, echo cancellation \cite{echo_ai} and even underwater communications \cite{underwater_ai}. Yet, despite its apparent universality, we argue that this paradigm shift does not represent the most appropriate path for communication systems, where physical principles and analytical rigor remain indispensable. A closer look at the promises and pitfalls of AI is therefore essential to reflect on its true role in the design of nonlinear communication systems \cite{ai_pitfalls}. The future, therefore, should not be driven by uncritical replacement by AI for everything, but rather a careful integration of learning-based methods into the robust frameworks that communication engineers have developed over time.

In this work, we adopt the Hammerstein model, which is well suited to transmission systems characterized by a static nonlinearity (e.g., a power amplifier) followed by a linear time-invariant (LTI) channel (e.g., ARMA). Although the Hammerstein model may not theoretically capture all physical nonlinear communication channels, it effectively represents a broad range of them and serves as an insightful initial modeling framework. Furthermore, its structure offers a convenient balance between analytical tractability and physical relevance. This separation of nonlinear distortion and linear memory effects simplifies both theoretical analysis and practical algorithm design. The Hammerstein model can approximate a wide variety of nonlinear behaviors with relatively low computational complexity, making it a suitable foundation for developing and validating estimation and equalization techniques before extending them to more complex nonlinear models. Building upon this framework, this paper introduces the DCT-based neuron as a flexible modeling tool capable of accurately estimating the channel under this assumption. Once the general nonlinear channel is identified, we further discuss transceiver architectures that can best exploit this model to recover the desired signal.

\section{Efficient Nonlinear Modeling}
\label{sec:lms-dct}

In the design of a nonlinear signal processing framework, the model is as important as the strategy used to operate and control it. In many cases, the operational scheme must enable the system to learn or adapt to the underlying model from observed data. However, existing approaches often propose a model without fully considering the implications it imposes on the associated learning or adaptation algorithm.

Consider the domain $\mathcal{I}_N=[0,1,\dots,N-1]$ and a nonlinear function $f:\mathcal{I}_N\rightarrow\mathcal{I}_N$ acting on an input $x\in\mathcal{I}_N$. At this stage, $x$ represents a generic signal passing through an instantaneous memoryless nonlinear system. For instance, in the traditional case of a power amplifier, $f$ corresponds to an AM–AM distortion, where input amplitude affects output amplitude without altering phase. Extensions of this model are left for future work. 

A variety of approaches have been proposed for modeling nonlinear communication channels, ranging from black-box neural network architectures to physics-based models. 
Yet, for many practical communication systems, there is a strong need for models that strike a balance between generality, interpretability and computational tractability. This has motivated the widespread adoption of polynomial models \cite{GIANNAKIS2001533}, ranging from Volterra-series expansions \cite{predistortion_volterra} to orthogonal polynomials \cite{predistortion_orthogonal}, piecewise-linear approximations \cite{predistortion_piecewise} and other polynomial-based formulations \cite{predistortion_other, predistortion_other2, postdistortion_ofdm3, postdistortion_sparse}.
These models provide a natural extension of linear convolution to nonlinear settings to capture both instantaneous and memory-dependent effects. 
Crucially, they are flexible enough to describe a wide range of nonlinear hardware and channel effects, yet structured enough to enable efficient identification and mitigation in real time. As such, polynomial models have emerged as the workhorse of signal processing for modeling nonlinear communication channels. Nevertheless, these approaches seldom question the limitations of polynomial bases, both in terms of representational power and their impact on learning dynamics.

Next, we introduce and discuss limitations of polynomial models and then propose an alternative efficient nonlinear model.



\subsection{Polynomial models}

A typical polynomial approximation of order $Q$ is given by
\begin{equation}
    f(x)=\sum_{q=1}^QD_qx^q,
\end{equation}
where $x^q$ is the $q$th polynomial basis and $D_q$ its coefficient. Such approximations are local, stemming from a Taylor expansion at the origin. As $x$ moves away from zero, higher-order terms grow rapidly, leading to unbounded and often uncontrollable behavior.  

From a learning perspective, where adaptation consists of estimating the coefficients $D_q$ directly from data with gradient-based algorithms, polynomial models present serious challenges. Since polynomial functions grow exponentially with $Q$, high-degree terms produce large, unbounded gradients that often destabilize training. The lack of orthogonality among polynomial bases makes coefficient updates strongly coupled, so convergence becomes slow, initialization-dependent and sensitive to the chosen model order.
Furthermore, when the number of model parameters becomes comparable to the number of available data samples, the model tends to overfit. This manifests as oscillations between training points, where the model adapts too closely to noise, and poor generalization near the boundaries of the input domain.

\subsection{The DCT model}

In signal processing, the \gls{DFT} is a natural tool to represent discrete functions as sums of complex exponentials. Given a set of $N$ observation pairs $\{x,f(x)\}$, the DFT coefficients are obtained as
\begin{equation}
    F_k=\frac{1}{N}\sum_{x=0}^{N-1} f(x)e^{-j\frac{2\pi k x}{N}},
    \label{eq:DFT}
\end{equation}
where $N$ is the resolution of the DFT and $F_k\in\mathbb{C}$. The inverse DFT allows to reconstruct the function from the DFT coefficients as
\begin{equation}
    f(x)=\frac{1}{N}\sum_{k=0}^{N-1} F_ke^{j\frac{2\pi k x}{N}},
    \label{eq:iDFT}
\end{equation}

The asymptotic behavior of the DFT coefficients is closely tied to the smoothness of the original signal: if a function $f$ has $p$ continuous derivatives, its coefficients decay at a rate $\mathcal{O}(1/k^{p+1})$. This decay concentrates most of the energy in low-frequency terms, which is why Fourier-based methods are so effective for compression and approximation.

However, the DFT implicitly assumes that the function extends periodically beyond its domain $\mathcal{I}_N$. This periodicity creates discontinuities at the boundaries, even when the function itself is smooth within $\mathcal{I}_N$. The resulting jumps introduce spurious high-frequency transitions, forcing the DFT to rely on large coefficients at high $k$ to compensate for artifacts that are not intrinsic to the signal.

One possible solution to overcome this issue is to construct an even extension of the signal:
\begin{equation}
    \tilde{f}(x)=
    \begin{cases}
      f(x) & \text{if}\ 0 \leq n< N \\
      f(2N-x) & \text{if}\ N \leq n< 2N
    \end{cases}\quad
\end{equation}

When the DFT is applied to $\tilde{f}$, the even extension ensures that the periodic extension, implicitly assumed by the DFT, becomes smooth and continuous. As a result, the high-frequency coefficients that would otherwise be necessary to represent discontinuities in $f$ are no longer required. In fact, the DFT of $\tilde{f}$ simplifies to
\begin{align}
        {F}_k &= \frac{1}{N}\sum_{x=0}^{2N-1} \tilde{f}(x)\cos\left(\frac{2\pi k x}{N}\right) - \frac{j}{N}\sum_{n=0}^{2N-1}\tilde{f}(x)\sin\left(\frac{2\pi k x}{N}\right)\nonumber\\
        &=
        \frac{1}{N}\sum_{x=0}^{2N-1} \tilde{f}(x)\cos\left(\frac{2\pi k x}{N}\right),
        \label{eq:dft_to_dct}
\end{align}
where the imaginary terms vanish due to the even symmetry of the extension. This expression is equivalent, up to a scaling factor, to computing the DCT of the original function $f$:
\begin{equation}
F_{k} = \beta_k \sum_{x=0}^{N-1} f(x)\cos\left( \frac{\pi k(2x+1)}{2N} \right),
\label{eq:DCT}
\end{equation}
where $\beta_{0}=1/\sqrt{N}$ and $\beta_{k}=\sqrt{2/N}$ for $k>0$. The output values $F_k\in\mathbb{R}$ are called the DCT coefficients. Accordingly, the inverse DCT (iDCT) is defined as
\begin{equation}
f(x) = \sum_{k=0}^{N-1} \beta_kF_k \cos\left(\frac{\pi k(2x+1)}{2N}\right)
\label{eq:iDCT}
\end{equation}

\begin{figure}[t]
    \centering
    \begin{subfigure}[b]{0.47\columnwidth}
        \centering
        \includegraphics[width=\columnwidth]{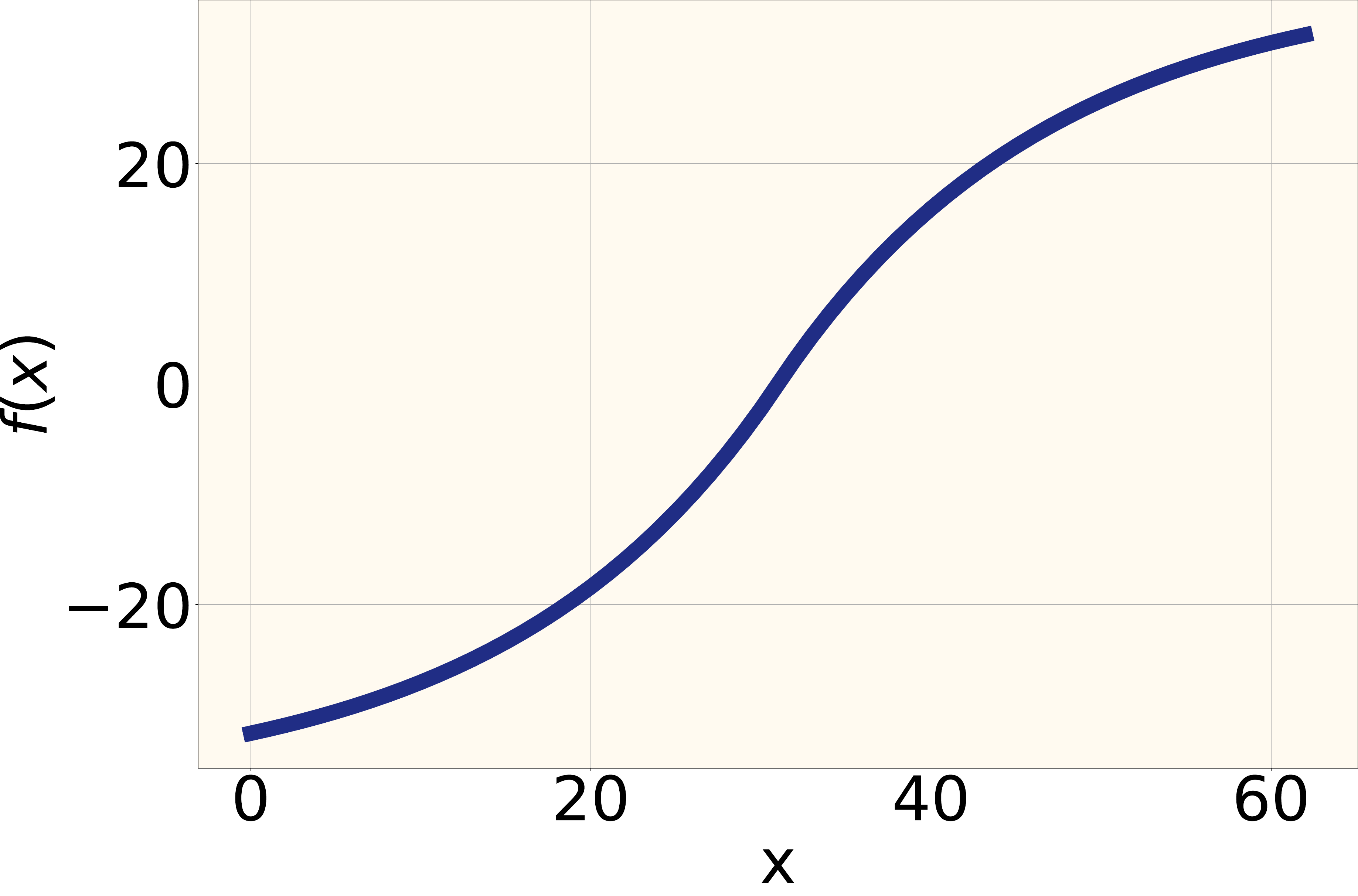}
        \caption{Function.}
        \label{fig:function_example}
    \end{subfigure}
    \hfill
    \begin{subfigure}[b]{0.47\columnwidth}
        \centering
        \includegraphics[width=\columnwidth]{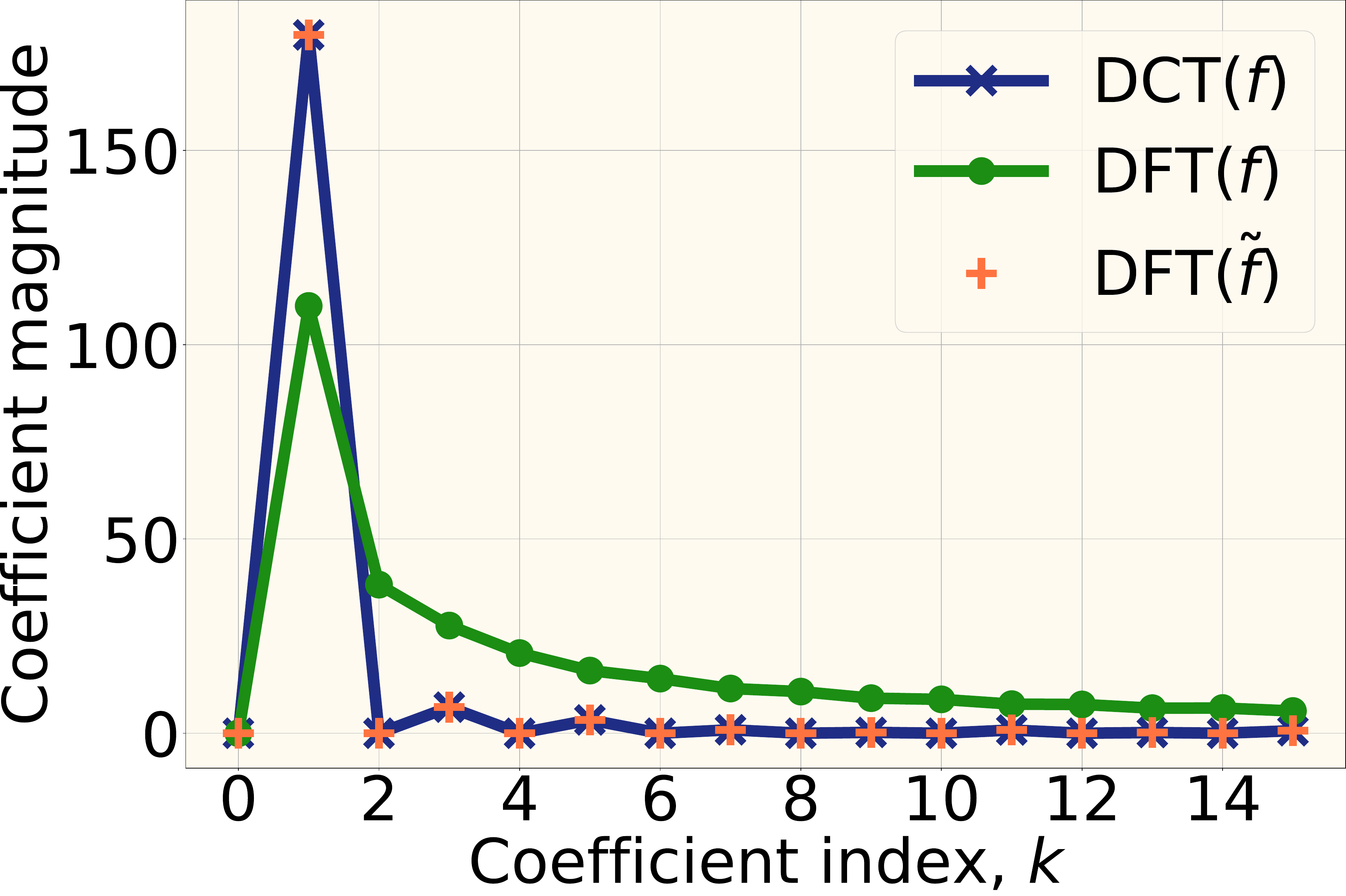}
        \caption{Coefficients.}
        \label{fig:DCT_DFT_coeffs}
    \end{subfigure}
    \caption{Illustration of the relationship between the DFT and DCT for the sigmoid function representation.}
    \label{fig:example_dct_dft}
\end{figure}

Thus, the DCT can be understood as a DFT applied to an even-extended version of the original signal. In other words, the DCT is essentially the DFT preceded by a preprocessing step that enforces even symmetry and resolves the boundary discontinuity problem.
Figure \ref{fig:example_dct_dft} illustrates the sigmoid function alongside the coefficients of the DFT, DCT and DFT for the symmetrically extended function. 

Another key advantage of the DCT is that the cosine functions form an orthogonal base. Orthogonality guarantees that each coefficient captures a distinct and independent contribution of the basis functions. As a result, changing the model order $Q$ (by adding or removing coefficients) does not alter the values of the existing ones. This property allows us to express the \gls{MSE} associated with approximating the function using $Q$ DCT coefficients as
\begin{equation}
    \text{MSE}=\mathbb{E}\left\{\left(f(x)-\hat{f}(x)\right)^2\right\}=\sum_{k=Q}^{N-1}F_k^2,
    \label{eq:mse_dct}
\end{equation}
where $\hat{f}(x)$ denotes the truncated $Q$-term version of \eqref{eq:iDCT}, and the expectation is taken with respect to $x$. In this form, the MSE is simply the energy of the discarded coefficients. Because the DCT naturally compacts most of the signal energy into the lowest-order terms, only a small number of coefficients is needed to achieve accurate approximations.

The DCT is one of the most influential mathematical tools in modern digital technology. Introduced in 1974 by Nasir Ahmed, together with T. Natarajan and K. R. Rao, the DCT became the cornerstone of image and video compression standards \cite{JPEG_standard, ITU-T_H264_V15}. It is at the heart of formats such as JPEG for images and MPEG for video technologies that made it possible to store, transmit and share digital media efficiently across the globe.
The impact of the DCT is difficult to overstate: every time we watch a streaming video, view a digital photograph or share multimedia on the internet, the DCT is working behind the scenes. Its ability to compact energy into a small number of coefficients makes it remarkably effective for compression, while still preserving perceptual quality. This balance of efficiency and fidelity is why the DCT has been called one of the most important innovations in the history of digital media.


\subsection{Learning the DCT coefficients}

We now propose to learn the DCT coefficients with an stochastic gradient descent algorithm, specifically, the \gls{LMS} algorithm \cite{Widrow_LMS}. Supervised approaches are particularly relevant when the analytical form of the target function $f$ is unavailable and the system must learn to approximate an unknown nonlinear mapping from observations. We show that, unlike polynomial models, the DCT makes nonlinear function approximation more efficient and analytically controllable. Further details of this method are discussed in \cite{LMS_DCT}.

\begin{figure}[t]
    \centering
    \includegraphics[width=0.8\columnwidth]{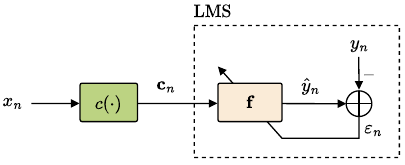}
    \caption{DCT-based neuron, where the LMS trains the coefficients.}
    \label{fig:LMS_DCT}
\end{figure}

We consider a set of scalar input-output pairs $\{x_n,y_n\}$, where $n$ is the sample index, $x_n\in\mathcal{I}_N$ and $y_n=f(x_n)$. As illustrated in Figure \ref{fig:LMS_DCT}, the scalar input $x_n$ is expanded into $Q$ cosine features with a nonlinear function $c$:
\begin{equation}
\mathbf{c}_n=c(x_n)=
\left[
\cos_1(x_n),\cos_2(x_n), \dots, \cos_Q(x_n)
\right]^T,
\label{eq:input_dct_lms}
\end{equation}
in which the cosine terms correspond to
\begin{equation}
\cos_q(x) = \cos\left(\frac{\pi(2q-1)(2x+1)}{2N}\right)
\label{eq:cos_def_new}
\end{equation}

The signal $\mathbf{c}_n$ is fed to the LMS algorithm and processed through a \gls{FIR} filter defined by a coefficient vector $\mathbf{f}\in\mathbb{R}^Q$. The output signal is
\begin{equation}
\hat{y}_n = \mathbf{f}^T \mathbf{c}_n = \sum_{q=1}^Q F_q \cos_q(x_n)=\hat{f}(x),
\end{equation}
which represents an approximation of the target function $f$ using $Q$ DCT coefficients. Finally, the LMS adapts the DCT coefficients using the error signal,
\begin{equation}
    \varepsilon_n=y_n-\hat{y}_n,
\end{equation}
and the following update rule,
\begin{equation}
    \mathbf{f}_{n+1}=\mathbf{f}_n+ \mu\varepsilon_n\mathbf{c}_n,
    \label{eq:rule_lms_gradient}
\end{equation}
where $\mu$ is the step-size.

In the standard LMS algorithm, the raw input $x_n$ is fed directly into the filter, resulting in a purely linear model. By contrast, the proposed approach uses the DCT-based neuron; that is, it introduces a nonlinearity through the function $c(.)$, while keeping adaptation linear with respect to the filter coefficients $\mathbf{f}$. This preserves the theoretical guarantees of LMS while extending its modeling capacity.

Trained using the LMS, the coefficients converge in average to 
\begin{equation}
    \mathbf{f}^*= \mathbf{R}_c^{-1}\mathbf{r}_{xc}=2\mathbf{r}_{xc},
    \label{eq:blocSol}
\end{equation}
where $\mathbf{r}_{xc}$ is the cross-correlation between the reference signal $x_n$ and the cosine vector $\mathbf{c}_n$. Since the DCT kernels are orthogonal and bounded, the autocorrelation reduces to a diagonal matrix with constant entries, $\mathbf{R}_c = \mathbb{E} \left[\mathbf{c}\mathbf{c}^T \right] =\frac{1}{2}\mathbf{I}$, and \eqref{eq:blocSol} reduces to $\mathbf{f}^*=2\mathbf{r}_{xc}$. This means that increasing the filter order ${Q}$ simply requires computing one additional correlation term of the vector $\mathbf{r}_{xc}$.

Furthermore, the input correlation matrix $\mathbf{R}_c$ is diagonal regardless of the statistics of $x_n$. Consequently, unlike the conventional LMS applied to a linear filter, where performance and stability bounds depend directly on the statistical properties of $x_n$, this dependency does not arise when the LMS is applied to the DCT-based neuron. This property significantly simplifies both filter design and step-size selection, leading to stable and predictable convergence behavior. 

Particularly, the step-size reduces to
\begin{equation}
    \mu=\frac{2\alpha}{\lambda_\text{max}}=
    4\alpha,
    \label{eq:step_size_LMS}
\end{equation}
where $\lambda_\text{max}$ is the maximum eigenvalue of $\mathbf{R}_c$ and $0<\alpha<1$ is a design parameter. The number of samples required to converge is
\begin{equation}
     T_{\kappa} \approx
     -\frac{\ln\left({\kappa}\right)}{2\alpha\lambda_\text{min}/\lambda_\text{max}}
     =-\frac{\ln\left({\kappa}\right)}{2\alpha},
    \label{eq:convergence_time}
\end{equation}
where $\ln(\cdot)$ is the natural logarithm and $\kappa$ indicates convergence when the residual error in the coefficients has reduced to $\kappa$. 

From \eqref{eq:step_size_LMS}, the step-size can be set without estimating input power, since the maximum eigenvalue of the DCT-transformed input matrix is equal to $1/2$. Moreover, because all eigenvalues of $\mathbf{R}_c$ are identical, the eigenvalue spread is exactly 1, ensuring the fastest possible convergence rate for LMS. Finally, note that the steady-state LMS misadjustment in the power error is proportional to $Q\alpha$. Consequently, as the number of coefficients increases, a larger number of training samples is required to achieve the same level of accuracy. Alternatively, maintaining a constant final misadjustment irrespective of ${Q}$ leads to a longer convergence time $T_{\kappa}$.

In summary, the DCT not only extends the modeling capabilities of a linear filter but also integrates seamlessly with the learning dynamics of the LMS algorithm. As a result, the DCT-based model effectively overcomes the instability and convergence issues typically associated with polynomial models. 

\begin{figure}[t]
    \centering
    \begin{subfigure}[b]{0.47\columnwidth}
        \centering
        \includegraphics[width=\columnwidth]{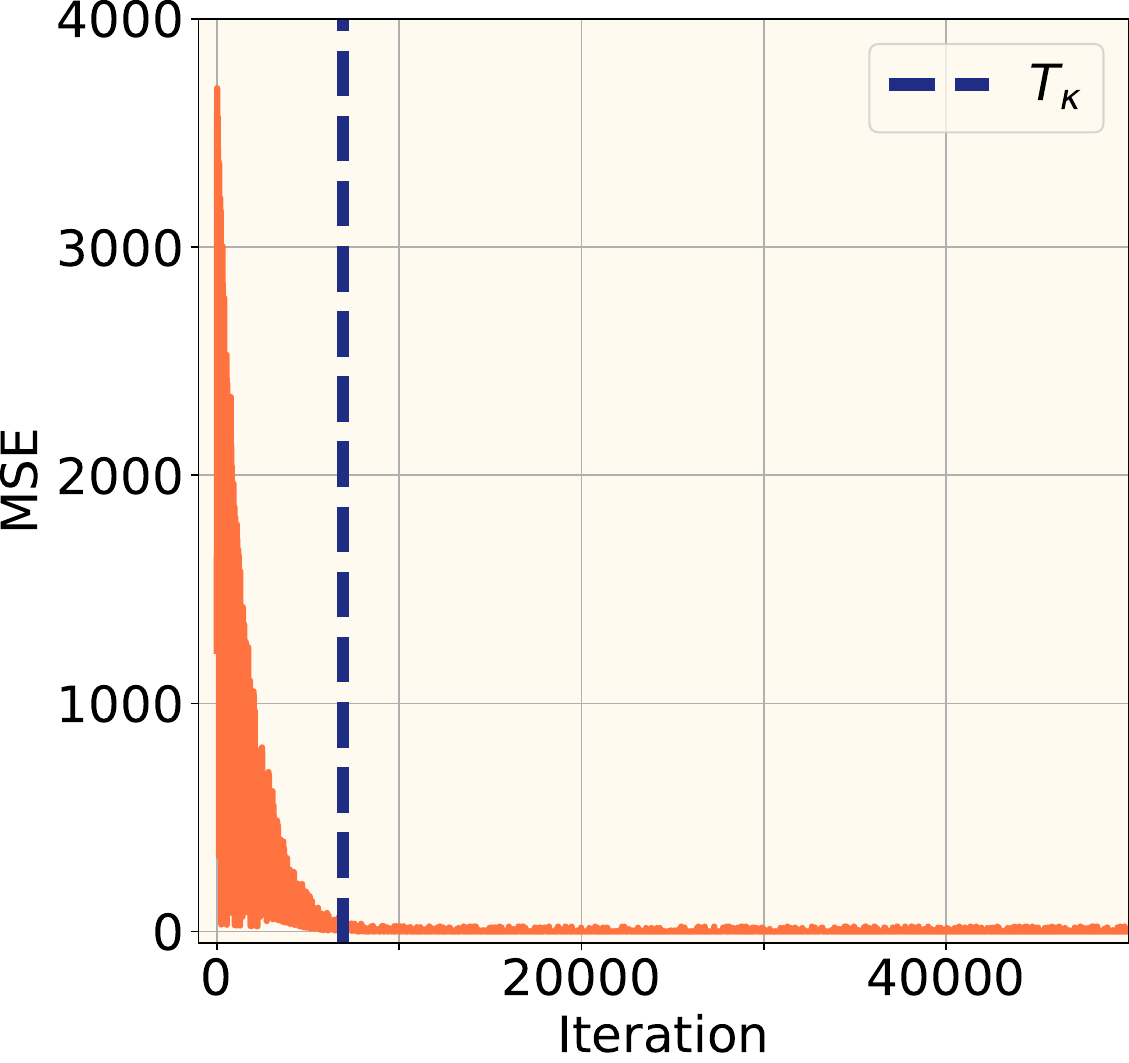}
        \caption{MSE loss and $T_{0.01}$.}
        \label{fig:LMS_sqrt_loss}
    \end{subfigure}
    \hfill
    \begin{subfigure}[b]{0.47\columnwidth}
        \centering
        \includegraphics[width=\columnwidth]{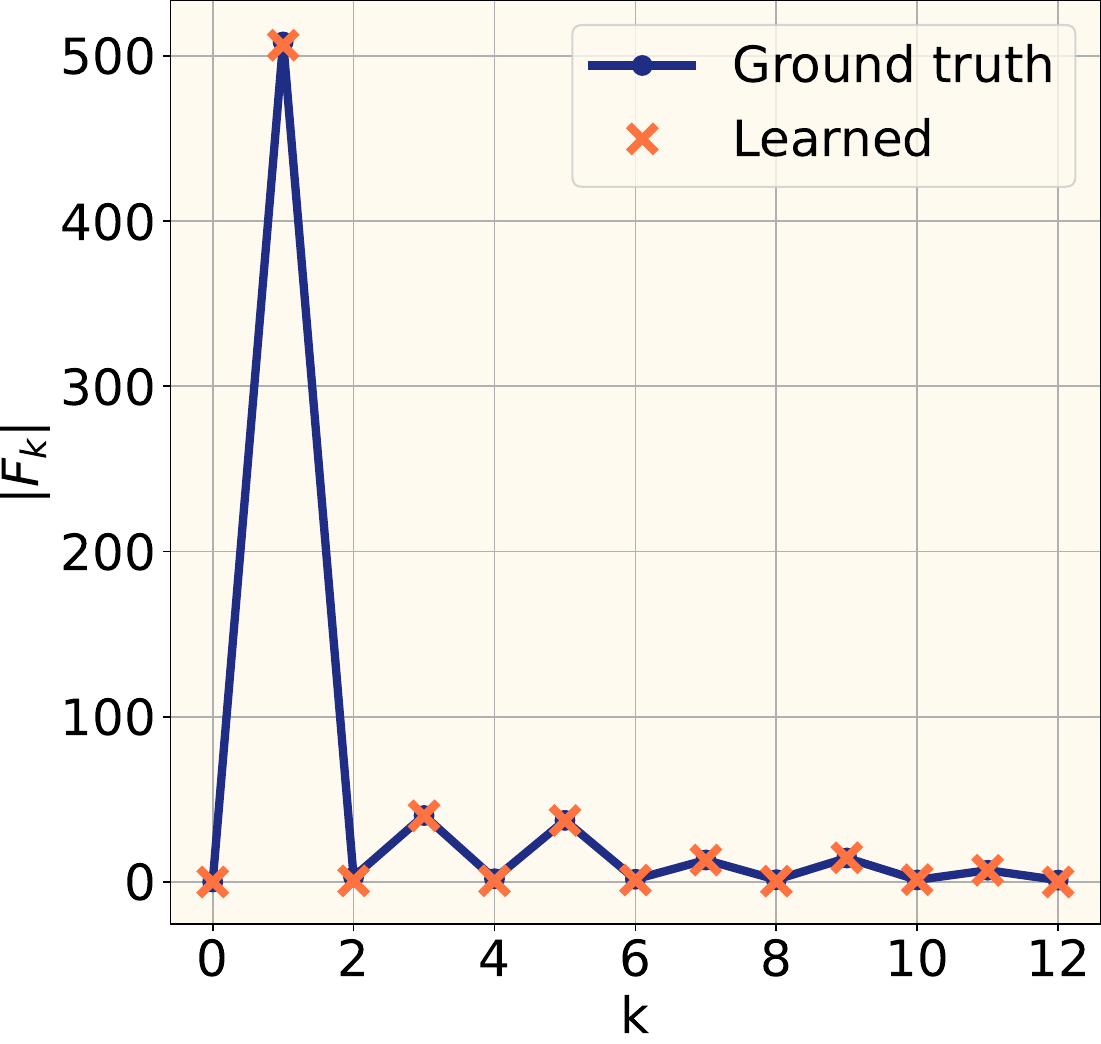}
        \caption{DCT coefficients.}
        \label{fig:LMS_sqrt_coeffs}
    \end{subfigure}
    \caption{Performance of the DCT-based LMS algorithm for the square root function.}
    \label{fig:example_lms}
\end{figure}

Figure \ref{fig:example_lms} illustrates the performance of the DCT-based LMS algorithm applied to learning the sigmoid function. The step-size is set to $\alpha=10^{-3}$ and the number of DCT coefficients is $Q=6$. This configuration has enough capacity to represent a wide variety of functions.
Using $\kappa=0.01$, the theoretical convergence time is $T_\kappa = 6.908$ samples, which closely matches the stabilization of the empirical MSE curve. Moreover, the learned DCT coefficients closely match the theoretical values, demonstrating the optimality of the DCT for function representation.

It is important to emphasize that 
being grounded in fundamental principles of function representation, the DCT yields a model that is both compact and well-controlled in terms of capacity, design and performance. 

\section{Nonlinear Channel Estimation in Flat Fading Channels}

The previous section discussed the adaptive design of the DCT-based neuron (or model). We now turn to its application in channel estimation within transceiver architectures. In particular, we focus on a canonical point-to-point communication scenario characterized by nonlinear AM–AM distortion and additive white Gaussian noise (AWGN).

Consider a baseband communication model, in which a real signal $x_n$ propagates through a nonlinear communication channel. The signal first undergoes an instantaneous memoryless nonlinear transformation $f$, after which \gls{AWGN}, denoted by $w_n \sim \mathcal{N}(0, \sigma^2)$, is introduced. 
The received signal is
\begin{equation}
    r_n=f(x_n)+w_n
    \label{eq:received_signal_channel}
\end{equation}

This model is sufficiently general to capture the behavior of flat-fading channels, in which case the channel gain is a constant factor absorbed by $f$. The only assumption is that of a block-fading channel, where the channel remains constant throughout the training phase.

The earliest attempts to mitigate nonlinearities in communication systems relied on inversion techniques, where the goal is to learn a function $g$ that approximates the inverse of $f$. The motivation behind studying the inverse is that, in principle, it allows the channel effects to be reversed, enabling recovery of the original transmitted signal $x_n$. When implemented at the transmitter side, before additional channel effects, this approach is known as predistortion \cite{predistortion_volterra, predistortion_orthogonal, predistortion_piecewise, predistortion_other, predistortion_other2}; when performed at the receiver side, it is referred to as equalization. In any of these situations the channel must be estimated, which is precisely the focus of this paper. Channel estimation is done at the receiver-side, since the receiver has access to pilot sequences used for channel synchronization, which can also be leveraged for channel estimation. Focusing on the receiver not only simplifies the transmitter design but also enables estimation of the entire channel distortion, including nonlinearities with memory that may arise after the power amplifier (e.g., due to bias or coupling impedance networks), as well as tracking their potential time variations.

Once the channel has been learned at the receiver, nonlinear compensation may be performed either at the receiver itself or at the transmitter.  Whenever feasible, transmitter-side predistortion is preferable, as it operates in a noise-free domain and thus preserves signal integrity. This highlights an essential design principle: the placement of processing intelligence within the transceiver chain fundamentally shapes system performance. In linear channels, the classical matched filter pair provides the optimal transmitter–receiver configuration under AWGN and ISI conditions \cite{Sklar2001}, while in MIMO systems, this concept extends naturally to joint transmit–receive optimization \cite{Palomar2003}.

For nonlinear channels, however, the notion of optimality remains an open question \cite{Oppen2010}. This work contends that in any nonlinear transceiver architecture, whether based on the proposed DCT model or on more advanced AI-driven designs, the learning of the channel should originate at the receiver, where the true, noise-affected signal is observed. Once acquired, this knowledge may be fed back to the transmitter for predistortion, or retained locally for adaptive detection, depending on system complexity, latency and power constraints.
Ultimately, this perspective reflects a broader insight: the intelligence of a communication system must be placed where information about distortion is richest, even if the correction is applied elsewhere.

Next, we study the two approaches: direct estimation, where the receiver learns the channel response itself, and the inverse estimation, where the receiver learns the inverse of the channel. Both strategies are formulated within the DCT-based framework and their performance are compared.

\subsection{Inverse Estimation}

In this inverse estimation setting, the receiver observes the received signal $r_n$ together with a known pilot symbol $x_n$, which serves as a reference for learning. Using these pairs, the receiver trains a parametric model $g$ such that
\begin{equation}
    \hat{x}_n=g(r_n)=g\left(f(x_n)+w_n\right)
    \label{eq:forward_processing}
\end{equation}

The instantaneous error is $\varepsilon_n = x_n - \hat{x}_n$, and in the absence of noise, the optimal mapping is $g=f^{-1}$. 

When $g$ is modeled using the DCT representation, the signal $r_n$ is processed by a \gls{FIR} filter and
the corresponding error signal becomes
\begin{equation}
    \varepsilon_n=x_n-\mathbf{f}^T\mathbf{c}_n,
\end{equation}
where $\mathbf{c}_n$ denotes the DCT projection of the received signal $r_n$. 
The optimal coefficients minimizing the MSE follow the standard solution formulated in \eqref{eq:blocSol}.
This estimation can also be implemented adaptively using the iterative update rule in \eqref{eq:rule_lms_gradient}. Also, note that the DCT implementation inherits all the benefits described in Section \ref{sec:lms-dct}.

It is important to highlight the fundamental challenges of this inverse estimation approach. From a representation perspective, the inverse of a nonlinear function may not exist; even when it does, representing it often requires a different model order. Specifically, if a function $f$ can be accurately represented with $Q$ coefficients, there is no guidance on how many parameters are needed to represent $f^{-1}$. 
Furthermore, as shown in \eqref{eq:forward_processing}, the nonlinear transformations distort the noise statistics, making the noise both non-additive and non-Gaussian. This poses a significant challenge, since many receiver components (e.g., symbol detectors, synchronization units, and timing recovery modules) are explicitly designed under the \gls{AWGN} assumption, and violating this assumption compromises their theoretical performance guarantees or requires a much higher \gls{SNR}. 



One might argue that using a more complex model for $g$ could lead to better performance. In this context, deep neural networks have also been proposed to learn inverse mappings in similar contexts, particularly in \gls{OFDM} system \cite{machine_learning_wireless}. While these methods can approximate highly nonlinear relations, they inherit the same conceptual limitations in the inversion of $f$. In fact, these issues may be exacerbated by the presence of multiple nonlinearities, further distorting the noise characteristics and complicating the receiver design.

An alternative approach to directly estimating the inverse consists of first estimating the direct channel response and subsequently inverting it. For univariate functions this effectively swaps the roles of $x$ and $f(x)$, transforming the original relation $y=f(x)$ into the reflected form $x=f^{-1}(y)$. Afterwards, the inverse \gls{DCT} in \eqref{eq:iDCT} is employed to approximate $f^{-1}$ as discussed in Section \ref{sec:lms-dct}. While the inverse function exhibits requires an increase in the number of coefficients $Q$, this inverse technique requires the same SNR as the direct estimation problem. As will be shown in the following section, this method is highly robust to noise and does not impose stricter SNR requirements.

\subsection{Direct Estimation}


\begin{figure}[t]
    \centering
    \begin{subfigure}[b]{\columnwidth}
        \centering
        \includegraphics[width=\columnwidth]{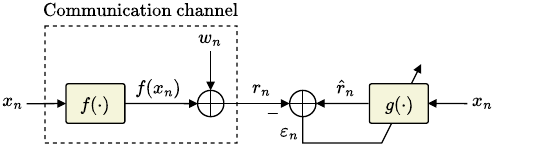}
        \caption{Direct channel estimation.}
        \label{fig:channel_compensation}
        \vspace{10pt}
    \end{subfigure}
    
    \hfill
    \begin{subfigure}[b]{\columnwidth}
        \centering
        \includegraphics[width=\columnwidth]{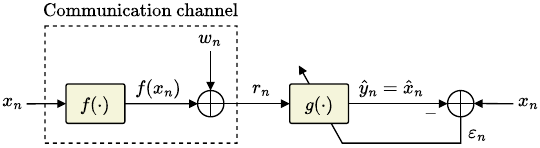}
        \caption{Inverse channel estimation.}
        \label{fig:channel_equalization}
    \end{subfigure}
    \caption{Supervised estimation of the direct and the inverse nonlinear channel in AWGN.}
    \label{fig:schemes_awgn}
\end{figure}


The fundamental principle of any optimal \gls{ML} receiver is that the communication channel should be learned, allowing the receiver to model the transformations experienced by the transmitted signal. Then, the receiver should select the most probable $\hat{x}_n$ given $r_n$. Under AWGN, and assuming the receiver knows $f$, the conditional probability density function of the received signal is:
\begin{equation}
    p(r_n \mid f(x_n)) = \frac{1}{\sqrt{2\pi\sigma^2}} \exp\left( -\frac{(r_n - {f}(x_n))^2}{2\sigma^2} \right)
\end{equation}
and $\hat{x}_n$ is obtained maximizing the log-likelihood, $\log p(r_n \mid f(x_n))$. This results in
\begin{equation}
    \hat{x}_n = \argmin_{x\in\mathcal{I}_N} \, \left(r_n - {f}(x)\right)^2,
    \label{eq:optimal_ml_channel}
\end{equation}
revealing that the ML solution corresponds to the $x_n$ that minimizes the distance between the observed data $r_n$ and the function output $f(x_n)$. Once an estimation of the channel is available, this approach is optimal with respect to the equalization one. Therefore, it is the best that can be implemented at the receiver.

Initially developed to address linear channel distortion, as described in \cite{lagunas_mdir}, subsequent works extended these ideas to nonlinear \gls{OFDM} channels \cite{cioffi_ml, postdistortion_ofdm2}. We revisit and build upon these earlier contributions, proposing the DCT model as a means to explicitly estimate the nonlinear component of the channel.


To perform direct channel estimation, the receiver uses a reference signal $x_n$ and applies a parametric model $g$ to approximate the channel output. By comparing $\hat{r}_n = g(x_n)$ with the actual received signal $r_n$, the corresponding error signal is
\begin{equation}
    \varepsilon_n=r_n-g(x_n)=
    f(x_n)-g(x_n)+w_n,
    \label{eq:error_compensation_channel}
\end{equation}
and minimizing the power of this error drives the function $g$ to approximate the true channel response $f$. 
When the DCT model is used, we can rewrite \eqref{eq:error_compensation_channel} as
\begin{equation}
    \varepsilon_n=r_n-\mathbf{f}^T\mathbf{c}_n,
\end{equation}
where $\mathbf{c}_n$ denotes the DCT projection of the pilot signal $x_n$. The optimal coefficients are
\begin{equation}
    \mathbf{f}^*= \mathbf{R}_c^{-1}\mathbf{r}_{rc}=2\mathbf{r}_{rc},
\end{equation}
where $\mathbf{r}_{rc}$ is the cross-correlation between the received signal $r_n$ and the cosine vector $\mathbf{c}_n$. This estimation can also be implemented adaptively, inheriting all the benefits of the DCT. 
Figure \ref{fig:schemes_awgn} illustrates the strategies for estimating both the inverse and direct channel responses.



It is fundamental to note that this estimation scheme does not process the noise component directly, preserving its original Gaussian distribution and additive nature. Hence, the MSE remains the optimal cost function to minimize, in accordance with the ML criterion. 

\subsection{Simulation Results}
\label{sec:results_awgn}

\begin{figure}[t]
    \centering
    \begin{subfigure}[b]{0.47\columnwidth}
        \centering
        \includegraphics[width=\columnwidth]{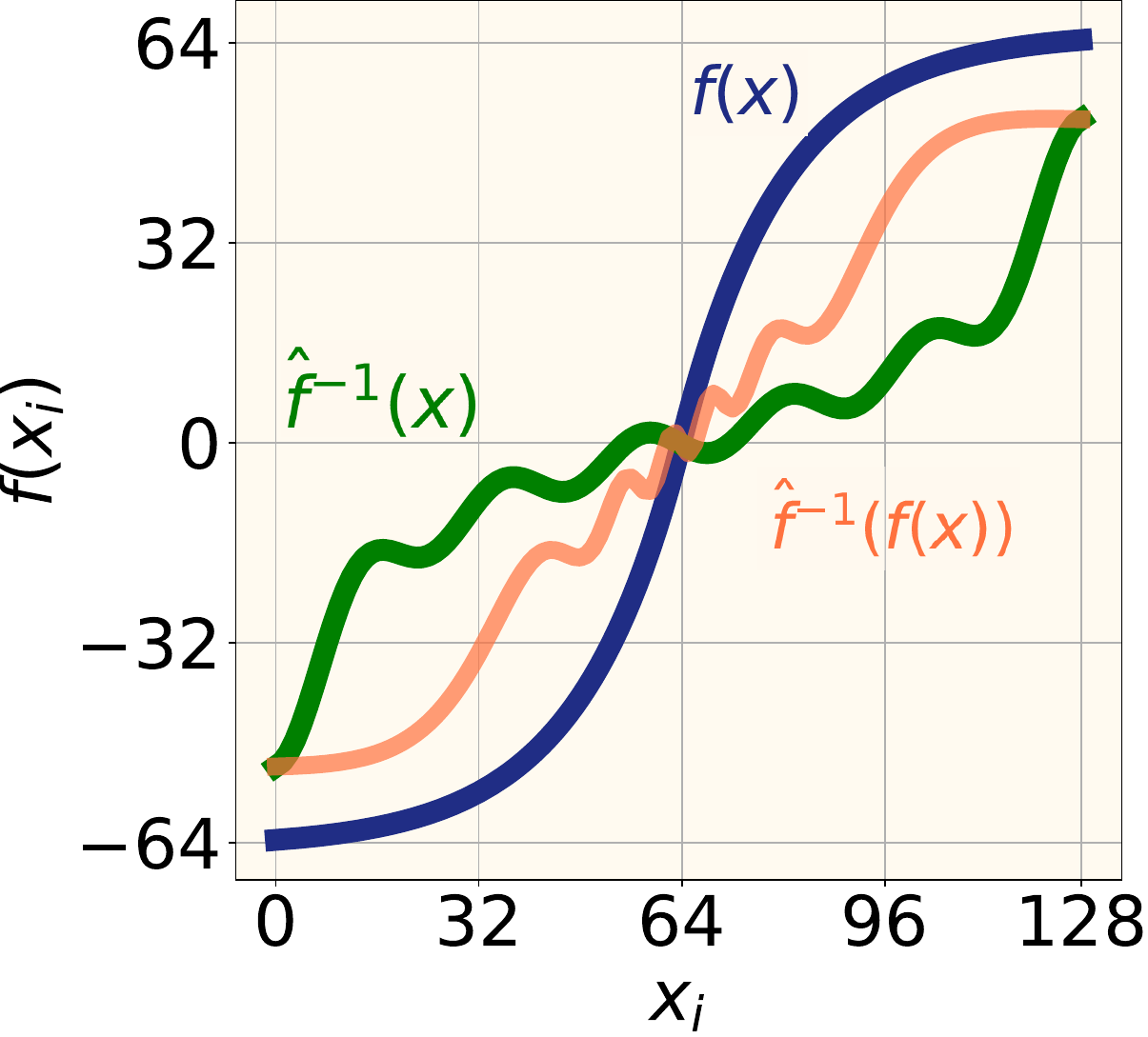}
        \caption{Inverse channel estimation.}
        \label{fig:inverse_channel_estimation}
    \end{subfigure}
    \hfill
    \begin{subfigure}[b]{0.47\columnwidth}
        \centering
        \includegraphics[width=\columnwidth]{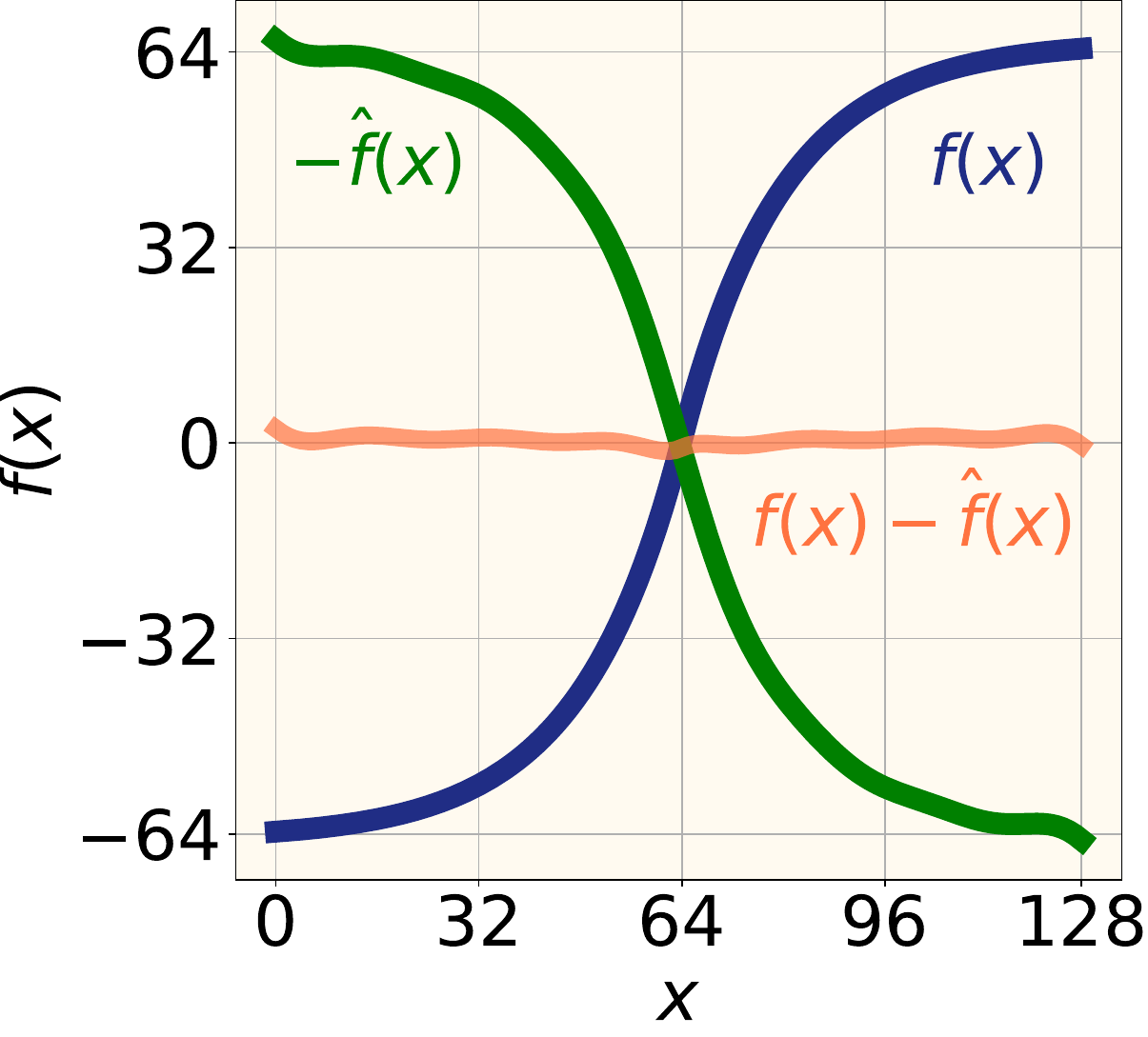}
        \caption{Direct channel estimation.}
        \label{fig:direct_channel_estimation}
    \end{subfigure}
    \caption{Comparison of inverse and direct channel estimation in an ideal noiseless channel ($\text{SNR}^{\text{pre}}=80$ dB) with $Q=6$ coefficients for both DCT models.}
    \label{fig:comparison_direct_inverse}
\end{figure}


In this subsection we aim to demonstrate the advantages of the direct channel estimation over inverse channel estimation in AWGN channels. 
All experiments are conducted using a dataset of 5,000 observations, uniformly sampled over $\mathcal{I}_N$ with $N = 128$. In all experiments, $\alpha = 1\times10^{-2}$.

We compare the proposed schemes with the same pre-detection \gls{SNR}, which is defined as the SNR at the input of the receiver:
\begin{equation}
    \text{SNR}^\text{pre}=\frac
    {\mathbb{E}\left\{ f(x_n)^2 \right\}}
    {\sigma^2}
\end{equation}

We begin by evaluating both methods in an idealized, noiseless communication scenario, that is, at $\text{SNR}^\text{pre} = 80$ dB. Figure \ref{fig:comparison_direct_inverse} presents the results for a nonlinear logarithmic compander. The figure depicts the original nonlinear function $f$ and the estimated function $g$. When estimating the inverse, we expect $g(x)\approx f^{-1}(x)$, whereas in the direct estimation, $g(x)\approx f(x)$. In both cases, the overall system response is shown in orange. Ideally, the inverse estimation should approximate the identity function, while the compensation response should correspond to the zero function. Note that in Figure \ref{fig:comparison_direct_inverse}(\subref{fig:direct_channel_estimation}) we plot $ -{g}(x)$ to improve visual clarity.

We note that the inverse approach exhibits poor performance, particularly near the boundaries of the input domain and within the linear region where noticeable rippling occurs. In contrast, the direct estimation method achieves excellent performance, producing a system response that is null over the entire domain. Using the adaptive algorithm, the \gls{MSE}, normalized with respect to the input power, reaches $1.48 \times 10^{-2}$ for the inverse estimation and $2.81\times10^{-4}$ for the direct estimation. Convergence is observed after 500 samples in the inverse estimation and about 300 samples in the direct one.

\begin{figure}[t]
    \centering
    \begin{subfigure}[t]{\columnwidth}
    \begin{subfigure}[b]{0.32\textwidth}
        \includegraphics[width=\textwidth]{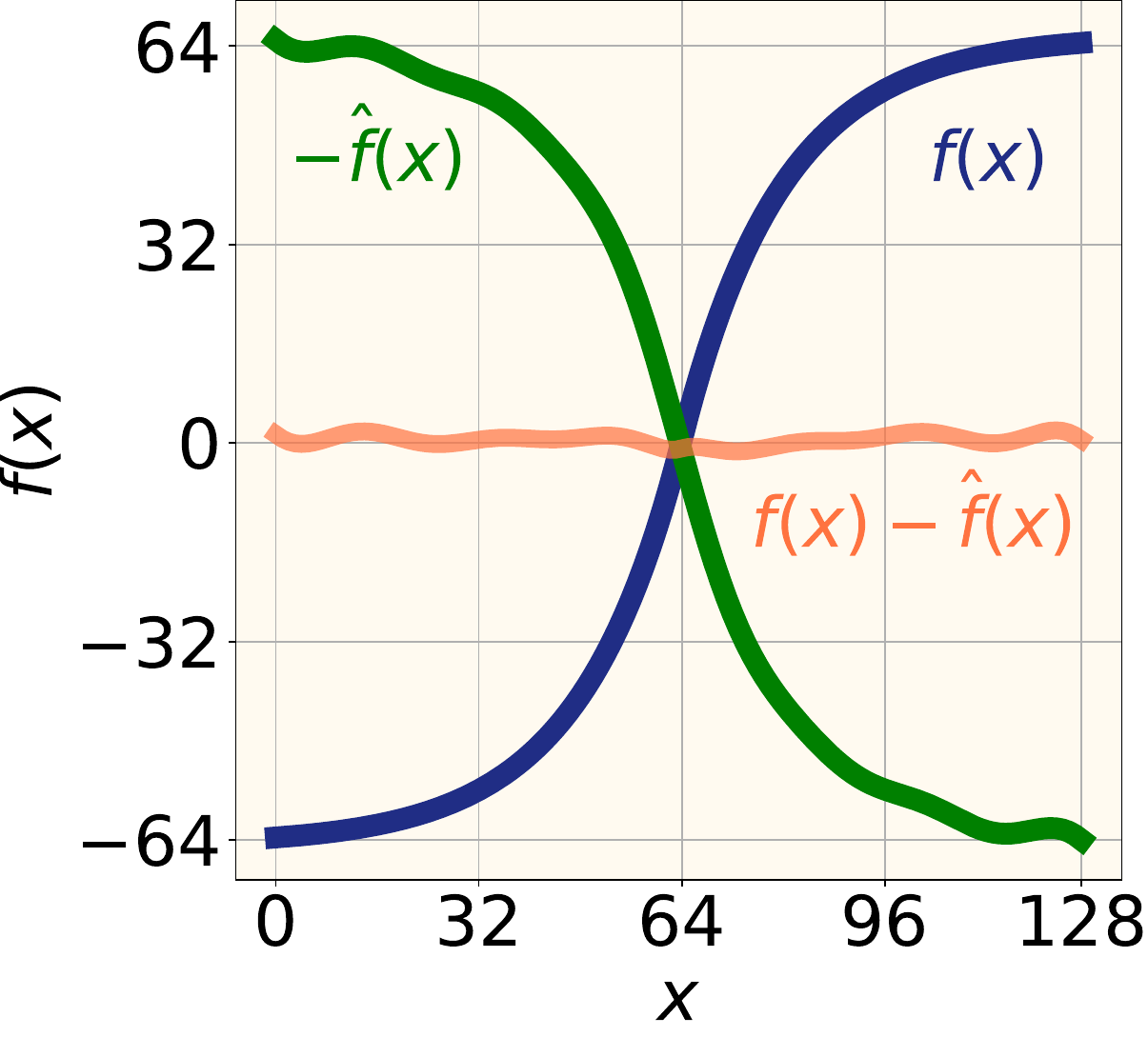}
    \end{subfigure}
    \hfill
    \begin{subfigure}[b]{0.32\textwidth}
        \includegraphics[width=\textwidth]{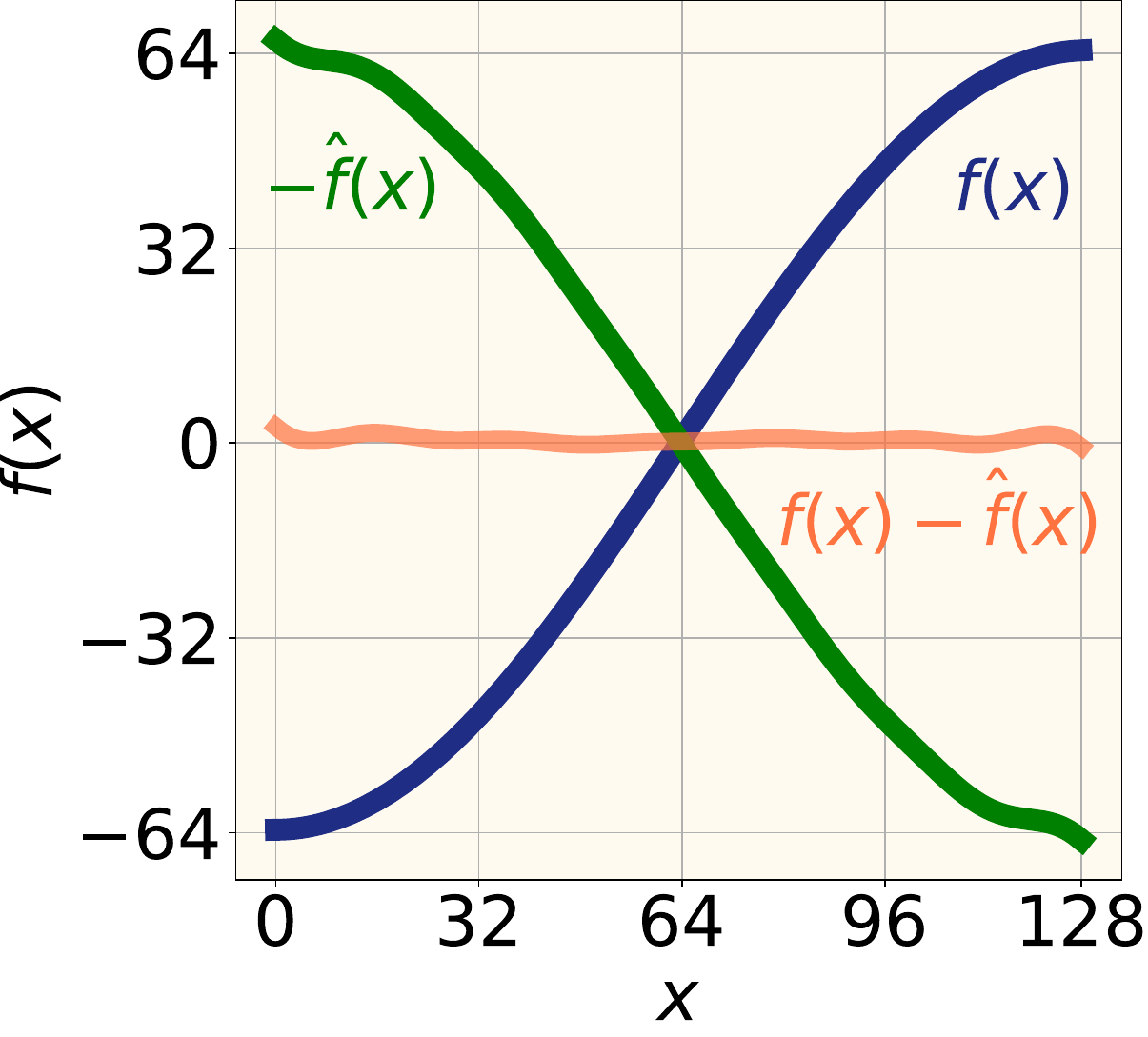}
    \end{subfigure}
    \hfill
    \begin{subfigure}[b]{0.32\textwidth}
        \includegraphics[width=\textwidth]{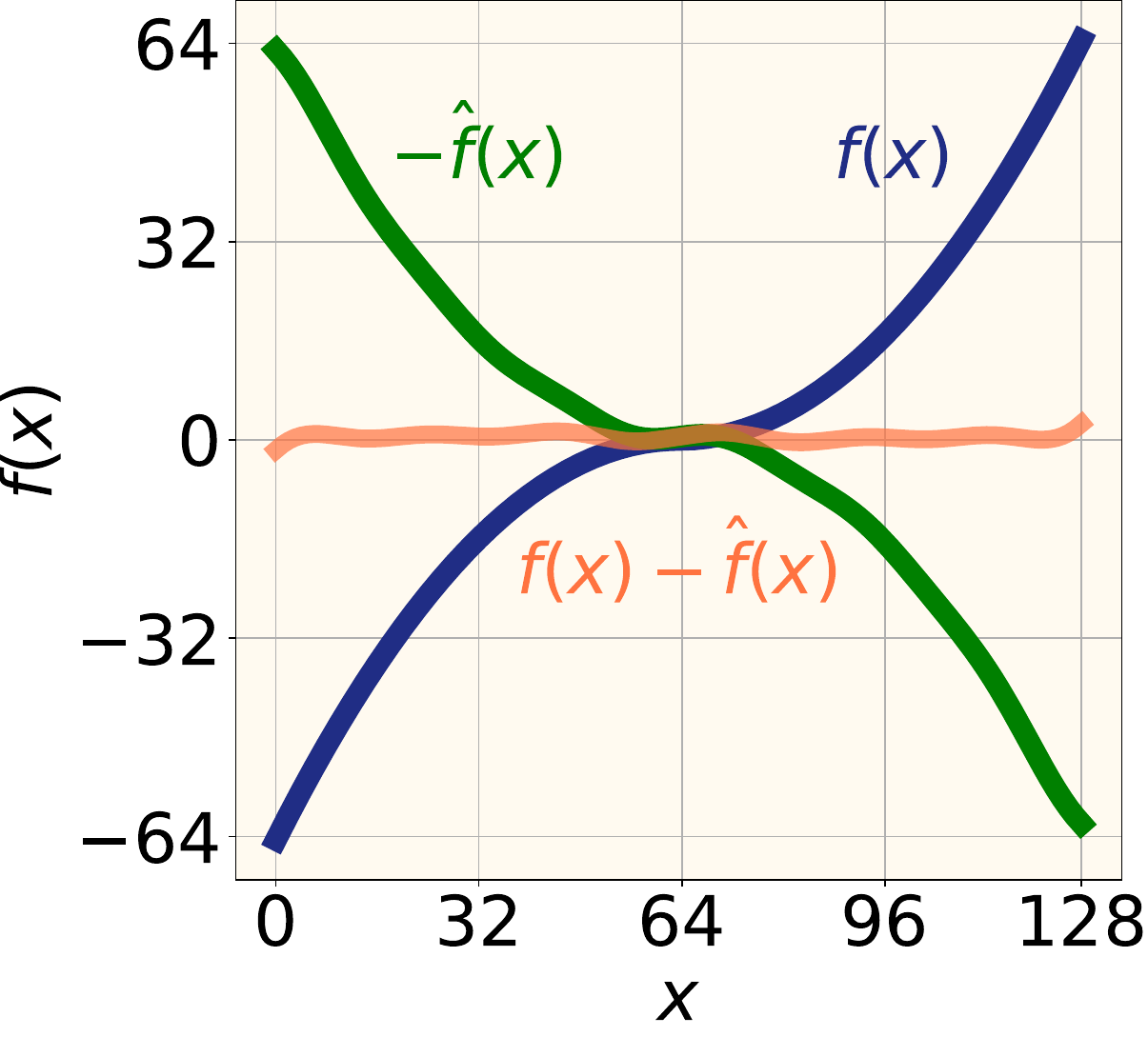}
    \end{subfigure}
    \subcaption{ $\text{SNR}^\text{pre}=0$ dB.}
    \label{fig:compensation_snr_0}
    \end{subfigure}\\[1.5em]
    
    \begin{subfigure}[t]{\columnwidth}
    \begin{subfigure}[b]{0.32\textwidth}
        \includegraphics[width=\textwidth]{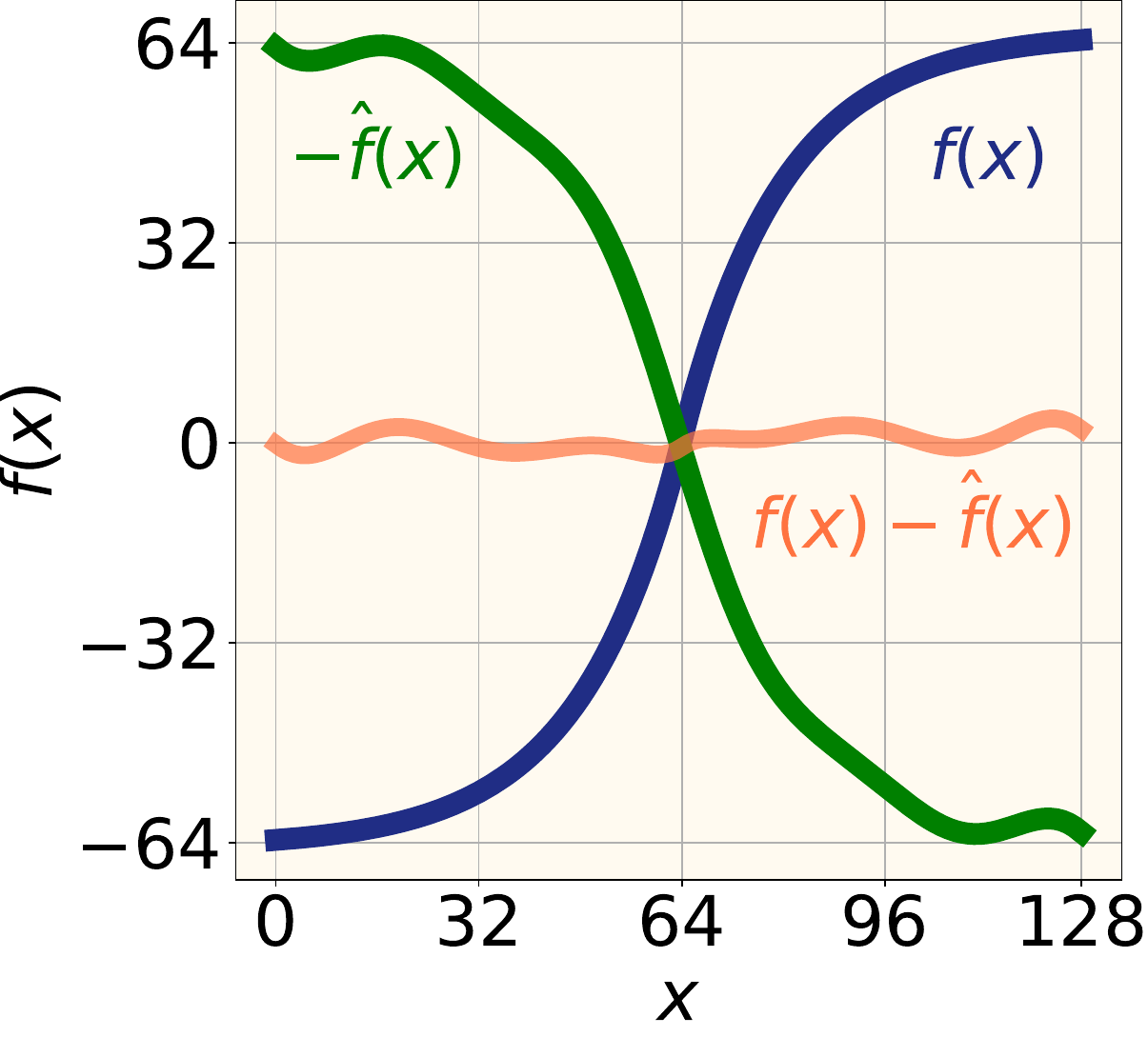}
    \end{subfigure}
    \hfill
    \begin{subfigure}[b]{0.32\textwidth}
        \includegraphics[width=\textwidth]{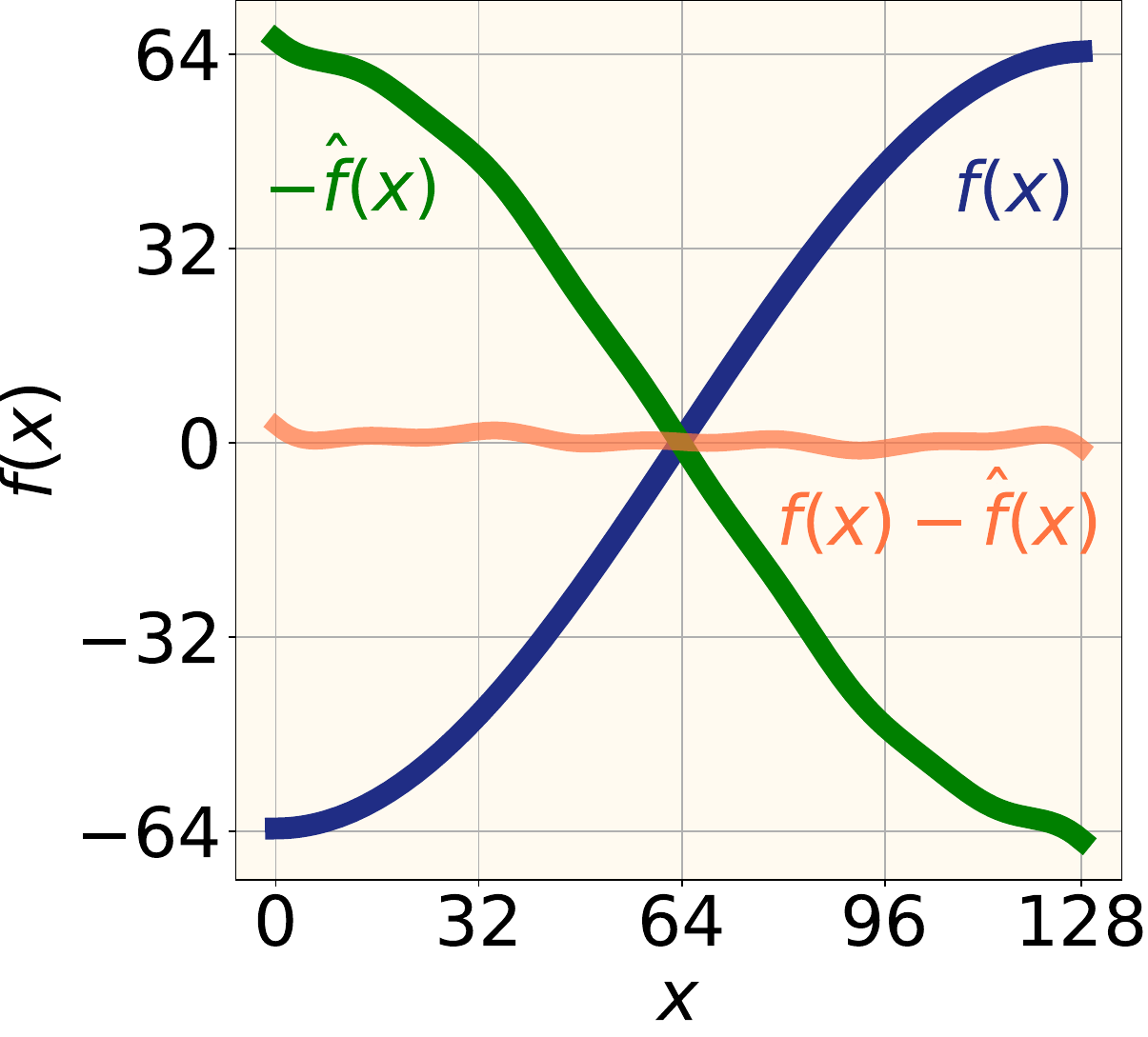}
    \end{subfigure}
    \hfill
    \begin{subfigure}[b]{0.32\textwidth}
        \includegraphics[width=\textwidth]{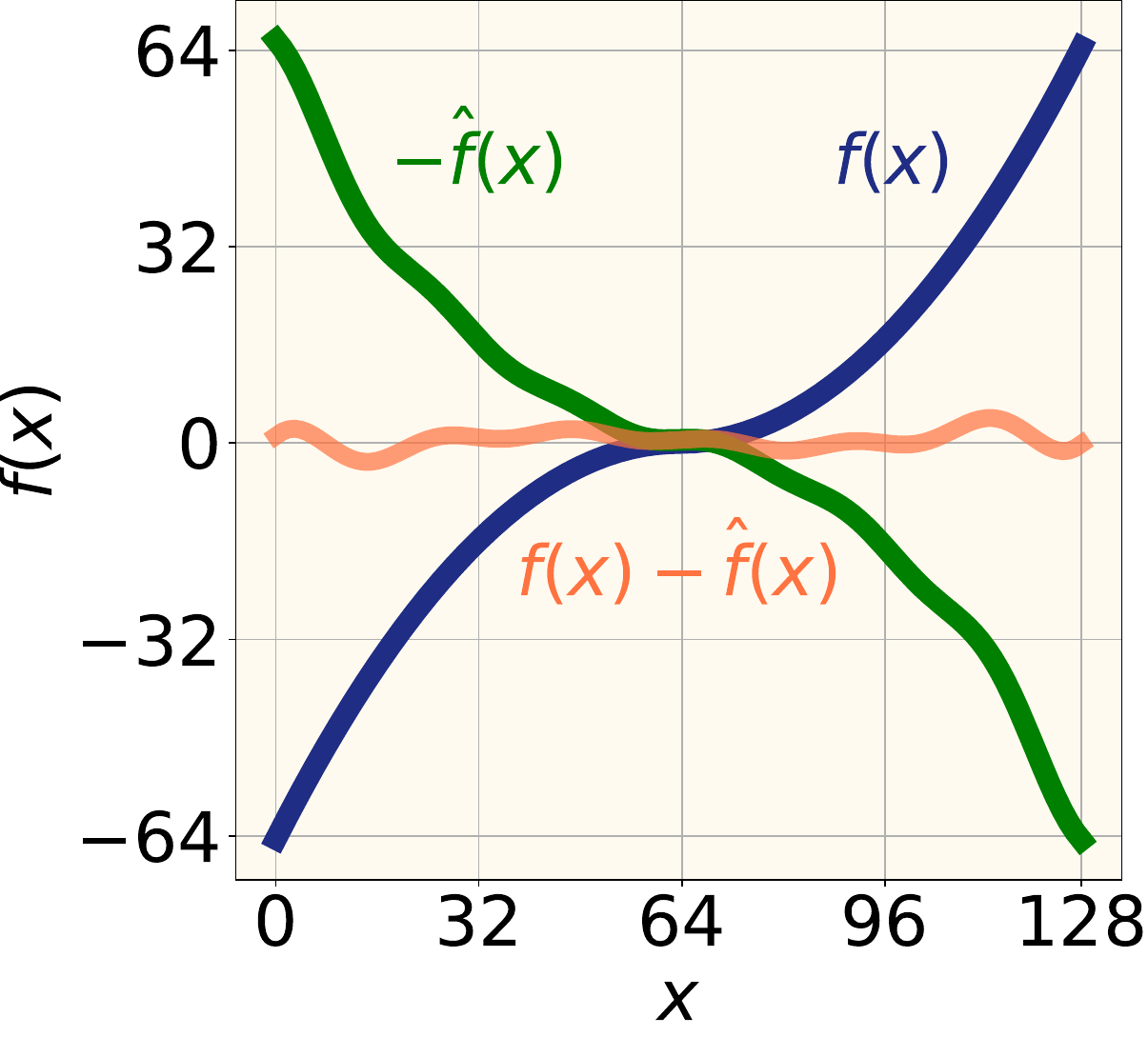}
    \end{subfigure}
    \subcaption{ $\text{SNR}^\text{pre}=-10$ dB.}
    \label{fig:compensation_snr_-10}
    \end{subfigure}

    \caption{Direct channel estimation in an AWGN channel at different $\text{SNR}^\text{pre}$ for different nonlinear functions $f$.}
    \label{fig:direct_estimation_snr}
\end{figure}

Note that the noise is not processed when performing the direct estimation and, therefore, it follows a Gaussian distribution. Consequently, using the \gls{MSE} to maximize the likelihood is the appropriate approach.
Figure \ref{fig:direct_estimation_snr} illustrates the performance of this method across different SNR regimes and nonlinear functions, specifically, sinusoidal and square nonlinearities. The direct channel estimation maintains good performance and exhibits a graceful degradation with increasing noise.
Conversely, the noise highly affects the inverse channel estimation and the method collapses at $-10$ dB. 

\begin{figure}[t]
    \centering
    \begin{subfigure}[b]{0.32\columnwidth}
        \centering
        \includegraphics[width=\columnwidth]{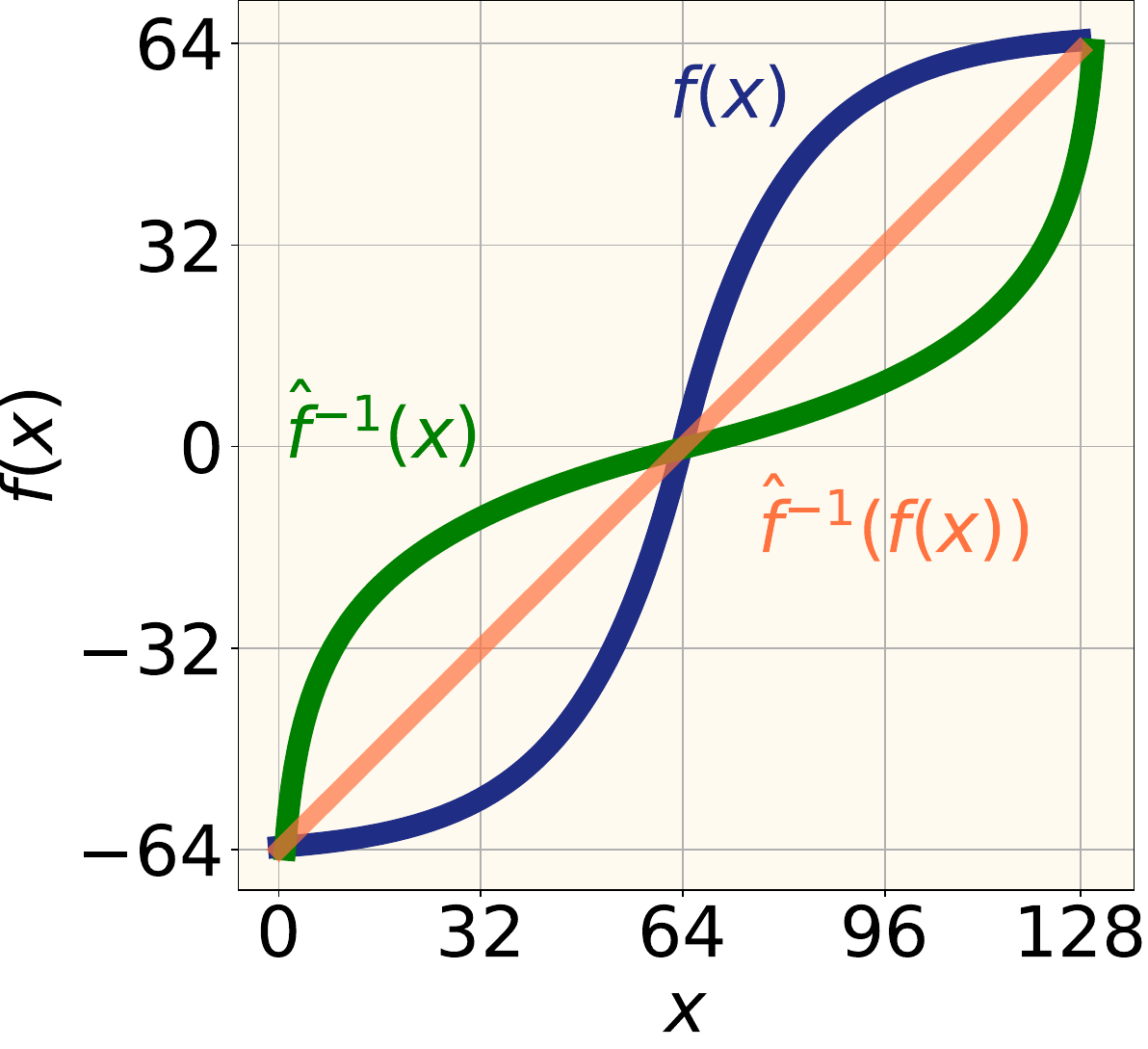}
        \caption{Compander.}
        \label{fig:inv_compander_snr_30}
    \end{subfigure}
    \hfill
    \begin{subfigure}[b]{0.32\columnwidth}
        \centering
        \includegraphics[width=\columnwidth]{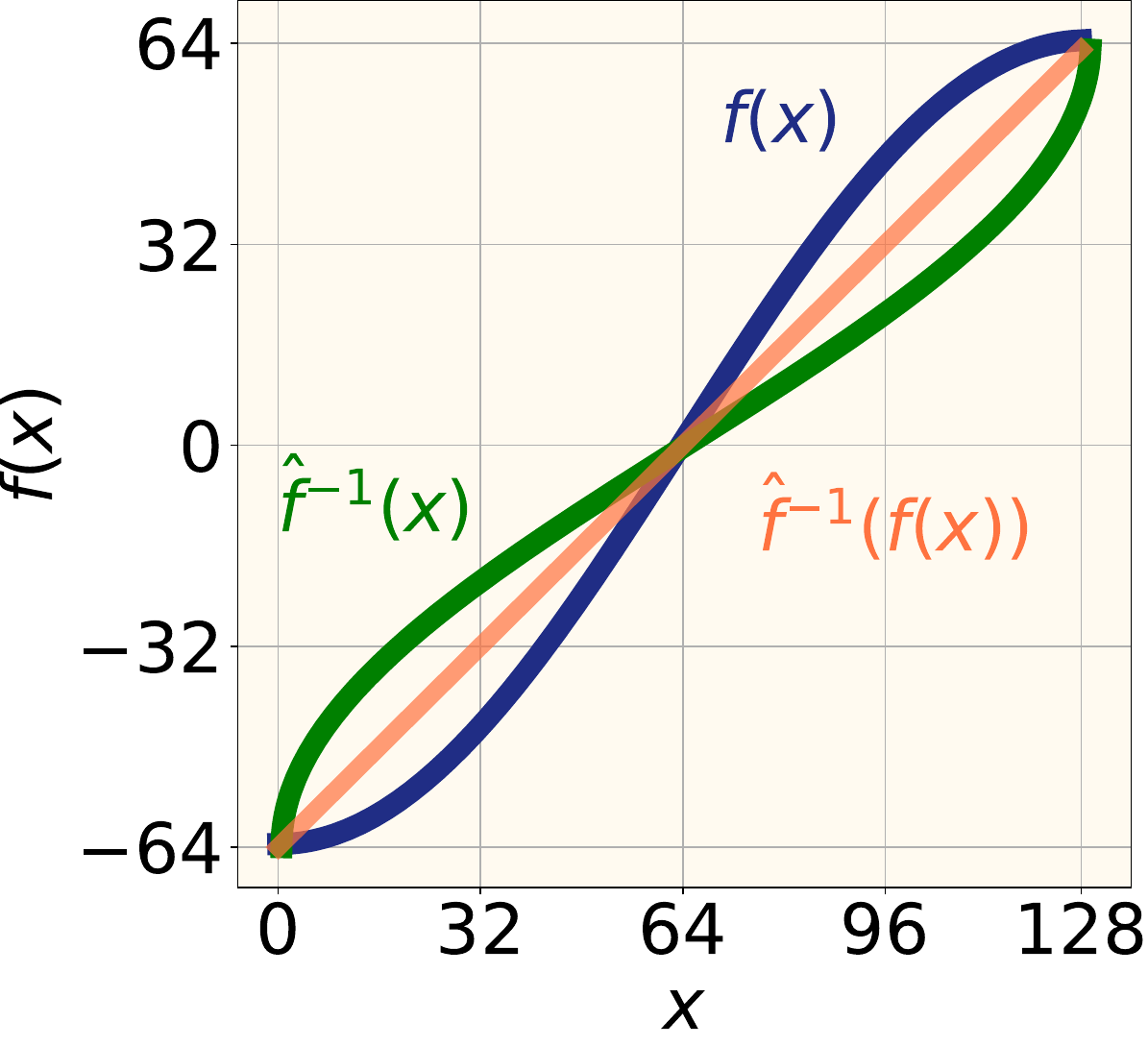}
        \caption{Sine.}
        \label{fig:inv_sin_snr_30}
    \end{subfigure}
    \hfill
    \begin{subfigure}[b]{0.32\columnwidth}
        \centering
        \includegraphics[width=\columnwidth]{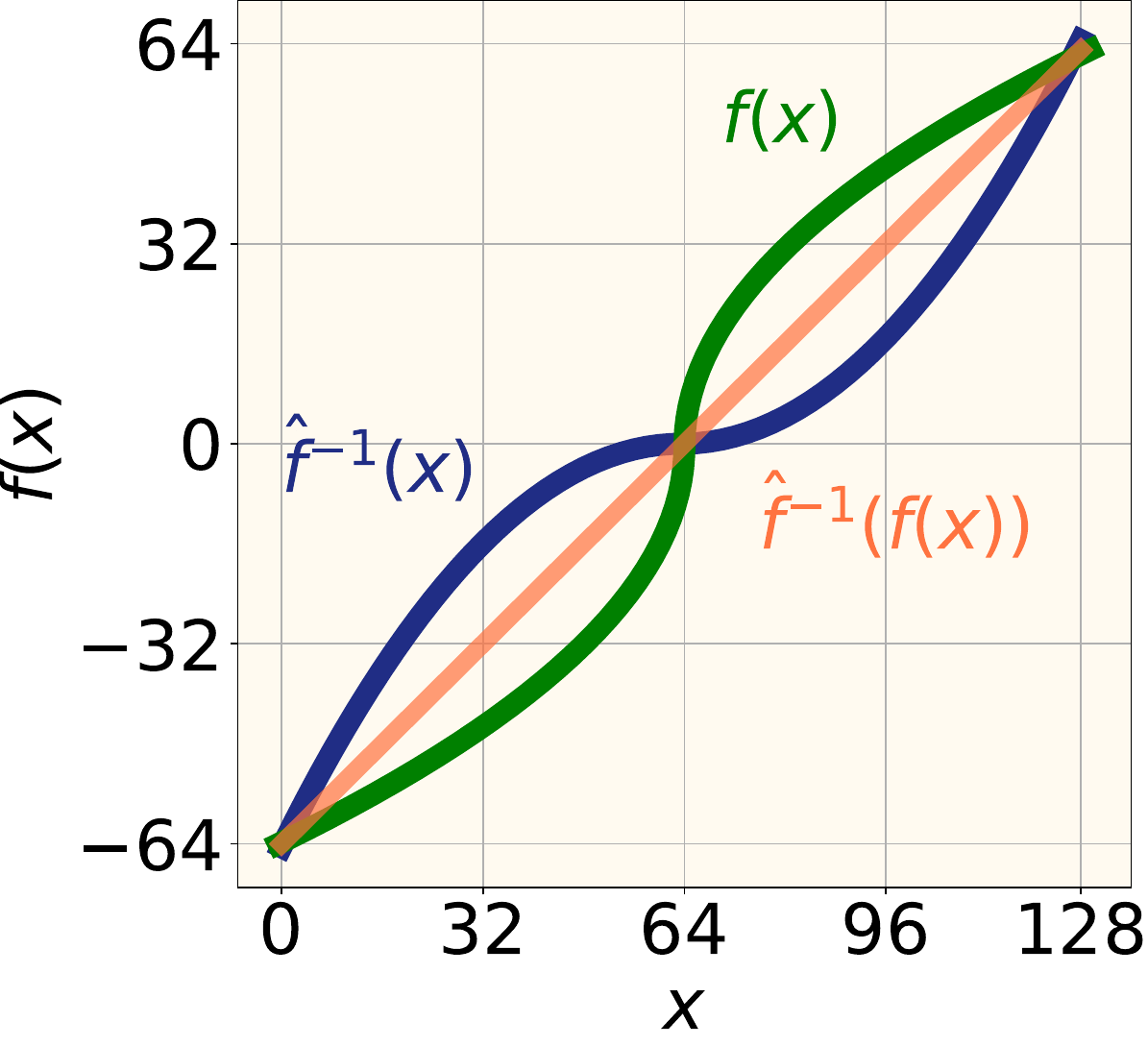}
        \caption{Square.}
        \label{fig:inv_square_snr_30}
    \end{subfigure}
    \caption{Inverse function obtained from the direct channel estimation using the DCT model at $\text{SNR}^{\text{pre}}=30$ dB. The DCT model employs $Q=6$ coefficients for the direct estimation and $Q=32$ for the inverse estimation.}
    \label{fig:inverse_from_direct}
\end{figure}

As previously explained, the inverse of $f$ can be directly computed from the direct channel estimation. Figure \ref{fig:inverse_from_direct} illustrates the inverse function estimation from the direct channel estimation in a scenario with $\text{SNR}^{\text{pre}}=30$ dB. The DCT model employs $Q=6$ coefficients for the direct channel estimation and $Q=32$ for the inverse. Due to the high quality of the direct estimation, the inverse obtained through this method remains robust to noise, in contrast to the inverse derived via the inverse channel estimation in Figure \ref{fig:comparison_direct_inverse}(\subref{fig:inverse_channel_estimation}).

\section{Direct Nonlinear Channel Estimation in Frequency-Selective Channels}

In the previous section, we addressed the general problem of estimating a noisy nonlinear ideal (or flat-fading) channel. We now extend this framework by incorporating a more general model of the communication channel that accounts for memory and propagation effects. In communication systems, frequency-selective channels naturally arise when the delay spread of the channel is larger than the inverse of the transmission bandwidth. In digital communications, this results in multiple delayed and attenuated versions of the signal symbol arriving at the receiver, introducing \gls{ISI} and complicating the detection process. 
In such cases, the symbol-by-symbol detection strategy designed previously is no longer adequate, and the receiver must perform sequence detection over a block of $M$ symbols, rather than treating each symbol independently.

We begin by examining the linear case, where the nonlinear function $f$ reduces to the identity function. Studying this simpler setting is essential, since it clarifies how a linear channel can be compensated. Once this foundation is established, we extend the analysis to the more general and challenging case of nonlinear channels.

\subsection{Linear Channel}

Considering an \gls{IIR} model for the communication channel, the received signal is
\begin{equation}
    r_n=\sum_{m=0}^{M-1}a_mx_{n-m}-
    \sum_{\ell=1}^{M-1}b_\ell r_{n-\ell}+w_n,
    \label{eq:channel_iir}
\end{equation}
where $\{a_m\}$ and $\{b_\ell\}$ represent the coefficients of the feedforward and feedback filters, respectively, and $M$ denotes the order, which is assumed to be of equal length for both filters. The transfer function of the channel in the $z$-domain is given by
\begin{equation}
    H(z)=\frac{A(z)}{B(z)}= \frac{\sum\limits_{m=0}^{M-1}a_mz^{-m}} {\sum\limits_{\ell=1}^{M-1}b_\ell z^{-\ell}}
\end{equation}

From an ML perspective, the receiver would ideally reconstruct the entire channel response and jointly detect the transmitted symbols. However, such an approach quickly becomes intractable as the number of channel taps grows. To address this, more structured receiver designs have been proposed that decompose the problem into simpler components, reducing complexity while preserving much of the performance.

To strike a balance between optimality and tractability, we adopt the Matched Desired Input Response (MDIR) \cite{lagunas_mdir}. Consider the equivalent channel representation in the $z$-domain:
\begin{equation}
    B(z)R(z)=A(z)X(z)+R(z)W(z),
\label{eq:received_z_domain}
\end{equation}
where $X(z)$, $R(z)$ and $W(z)$ denote the transforms of the transmitted signal, received signal and noise, respectively.
The receiver is designed to mirror this structure: The received signal $\mathbf{r}_n\in\mathbb{R}^M$ is processed by a forward equalizer $\hat{\mathbf{b}}\in\mathbb{R}^M$ which compensates for the feedback dynamics introduced by $B(z)$. In parallel, the reference signal $\mathbf{x}_n$ is passed through a backward equalizer $\hat{\mathbf{a}} \in \mathbb{R}^M$, which captures the multipath response $A(z)$. Both outputs are compared and, ideally, any residual difference can only be attributed to noise.

In the context of wireless communication, channels only exhibit multipath (i.e., $B(z) = 1$) and the MDIR does not require explicit compensation of $B(z)$, unless the receiver has an antenna array and carries out both spatial and temporal processing (see \cite{lagunas_mdir} for more details).

A key advantage of this design is that the forward equalizer acts solely as a linear combiner, preserving the noise statistics and avoiding additional coloration. While this distributed approach is not strictly ML optimal, it offers a very attractive trade-off by reducing the number of parameters to be estimated to just $2M$. This not only simplifies the implementation but also increases robustness in noisy environments, making MDIR a practical and effective alternative to full ML receiver.

The solution, fully derived in \cite{lagunas_mdir}, yields the following structure:
\begin{equation}
\begin{aligned}
    \hat{\mathbf{a}} &= \mathbf{R}_{rx}^T \hat{\mathbf{b}}, \\
    \mathbf{R}_r \hat{\mathbf{b}} &= (\lambda_\text{min}+1)\mathbf{R}_{rx}\mathbf{R}_{rx}^T \hat{\mathbf{b}},
\end{aligned}
\label{eq:mdir_solution}
\end{equation}
where $\mathbf{R}_{rx}$ is the cross-correlation matrix between $\mathbf{r}_n$ and $\mathbf{x}_n$.

\subsection{Nonlinear Channel}

The nonlinear channel is modeled as the cascade of an instantaneous nonlinearity followed by a linear system, corresponding to a Hammerstein structure. 
In this scenario, estimating the inverse response becomes even more challenging than in the linear case, since the inversion of such a composite model is generally non-trivial and often ill-posed.

\begin{figure}[t]
    \centering
    \includegraphics[width=\linewidth]{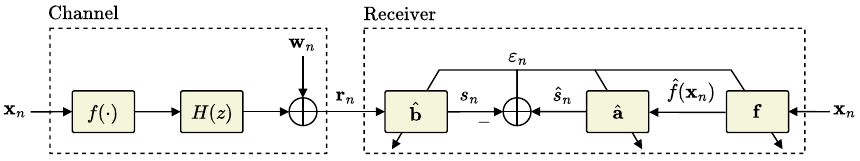}
    \caption{Receiver design to estimate a nonlinear frequency selective channel.} 
    \label{fig:nonlinear_channel_compensation}
\end{figure}

To address this challenge, we build upon the MDIR design and extend it to explicitly account for the nonlinear behavior of the channel. According to the ML criterion, the optimal receiver should mirror the structure of the channel itself and adopt the corresponding Hammerstein-like architecture. This entails replicating the nonlinear component at the receiver, ensuring that both the nonlinear and linear effects introduced by the channel are properly modeled.

As illustrated in Figure \ref{fig:nonlinear_channel_compensation}, the transmitted signal passes through an instantaneous nonlinearity, resulting in $f(\mathbf{x}_n)$. At the receiver, $f$ is replicated using the reference signal and a parametric model $g$, yielding $g(\mathbf{x}_n)$.
To handle the linear channel response $H(z)$, the MDIR equalizer processes the pair $\{ \mathbf{r}_n, g(\mathbf{x}_n) \}$ which are linearly related. The forward equalizer output, $s_n = \hat{\mathbf{b}}^T \mathbf{r}_n$, is then compared with the backward equalizer output, $\hat{s}_n = \hat{\mathbf{a}}^T g(\mathbf{x}_n)$, producing the error
\begin{equation}
    \varepsilon_n={s}_n-\hat{s}_n=
    \hat{\mathbf{b}}^T\mathbf{r}_n-
    \hat{\mathbf{a}}^T g(\mathbf{x}_n)
    \label{eq:error_nonlinear_channel_compensation}
\end{equation}

When the nonlinearity is represented using the DCT, then
\begin{equation}
\varepsilon_n=
\hat{\mathbf{b}}^T\mathbf{r}_n-
\hat{\mathbf{a}}^T\mathbf{C}_n\mathbf{f},
\label{eq:error_nonlinear_channel_compensation_DCT}
\end{equation}
where $\mathbf{C}_n \in \mathbb{R}^{M \times Q}$ is a matrix whose $m$th column corresponds to the DCT vector $\mathbf{c}_{n-m}$. More precisely, the $(m,q)$th entry of $\mathbf{C}_n$ is given by
\begin{equation}
\mathbf{C}_n^{(m,q)}=\cos\left(\frac{\pi q(2x_{n-m}-1)}{2N}\right)
\end{equation}

The MSE in \eqref{eq:error_nonlinear_channel_compensation_DCT} depends on the parameters $\hat{\mathbf{a}}$, $\hat{\mathbf{b}}$ and $\mathbf{f}$. While the coupling between $\hat{\mathbf{a}}$ and $\mathbf{f}$ renders the MSE jointly non-convex, it remains convex when each set of parameters is optimized independently. Therefore, we propose an alternating optimization approach where the DCT coefficients and MDIR parameters are updated iteratively. At iteration $n$:
\begin{itemize}
    \item The pair $\{\hat{\mathbf{a}},\hat{\mathbf{b}}\}$ is updated with \eqref{eq:mdir_solution}. Notice that the signal $\mathbf{x}_n$ has to be substituted by the new reference $g({\mathbf{x}_n})=\mathbf{C}_n\mathbf{f}$.
    \item To update the DCT coefficients, equation \eqref{eq:error_nonlinear_channel_compensation_DCT} can be reinterpreted as
    \begin{equation}
    \varepsilon_n=
    \hat{\mathbf{b}}^T\mathbf{r}_n-
    \mathbf{f}^T\mathbf{C}_n^T\hat{\mathbf{a}},
    \end{equation}
    which corresponds to the traditional LMS design where the reference signal is $\hat{\mathbf{b}}^T\mathbf{r}_n$, and the input vector is $\mathbf{C}_n^T\hat{\mathbf{a}}$. Then, the DCT coefficients can be updated as
    \begin{equation}
    \mathbf{f}_{n+1}=\mathbf{f}_{n}+4\alpha\varepsilon_n\mathbf{C}_n^T\hat{\mathbf{a}},
    \label{eq:update_rule_dct_channel}
    \end{equation}
\end{itemize}

It is worth highlighting that the benefits of the DCT model are also preserved under this formulation. 
The full learning procedure, including the forward equalizer, backward equalizer and DCT updates, is summarized in Algorithm \ref{alg: equalizer_and_dct}.



Once the channel is accurately estimated, the resulting model could be employed for symbol detection or waveform recovery, depending on the modulation format. In digital communication systems, the estimated channel can be incorporated into a Viterbi detector, which exploits the channel memory to achieve ML sequence detection. Alternatively, a \gls{DFE} provides a lower-complexity solution, which may also be extended to the restoration of analog waveforms. A detailed analysis of these detection strategies and their integration with the proposed nonlinear compensation framework is left for future work.

\begin{algorithm}[t]
\DontPrintSemicolon
\SetAlgoLined
\KwIn{ $\mathbf{r}_n$, $\mathbf{x}_n$, $\text{MSE}_\text{th}$ }
\vspace*{4 pt}
\KwOut{ $\hat{\mathbf{a}},\hat{\mathbf{b}}, \mathbf{f}$ }
\vspace*{4 pt}
Initialize $\mathbf{f}$ randomly\;
\vspace*{4pt}
 \While{\upshape MSE$\leq\text{MSE}_\text{th}$}{
    \vspace*{2pt}
    Compute $g(\mathbf{x}_n)$ with $\mathbf{f}$\;
    \vspace*{2pt}
    Update $\{\hat{\mathbf{a}},\hat{\mathbf{b}}\}$ with \eqref{eq:mdir_solution}\;
    \vspace*{4pt}
    Update $\mathbf{f}$ with \eqref{eq:update_rule_dct_channel}\;
 }
 \caption{Optimization of the equalizers and the DCT coefficients}
 \label{alg: equalizer_and_dct}
\end{algorithm}

\subsection{Simulation Results}

We now illustrate the performance of the nonlinear frequency selective channel estimation scheme. The performance of the MDIR for linear channels is not reproduced here and can be found in the original work \cite{lagunas_mdir}. 
We consider the same dataset used in Section \ref{sec:results_awgn} and maintain the same LMS configuration throughout. The channel order is $M=3$.

\begin{figure}[t]
    \centering
    \begin{subfigure}[t]{\columnwidth}
    \begin{subfigure}[b]{0.315\textwidth}
        \includegraphics[width=\textwidth]{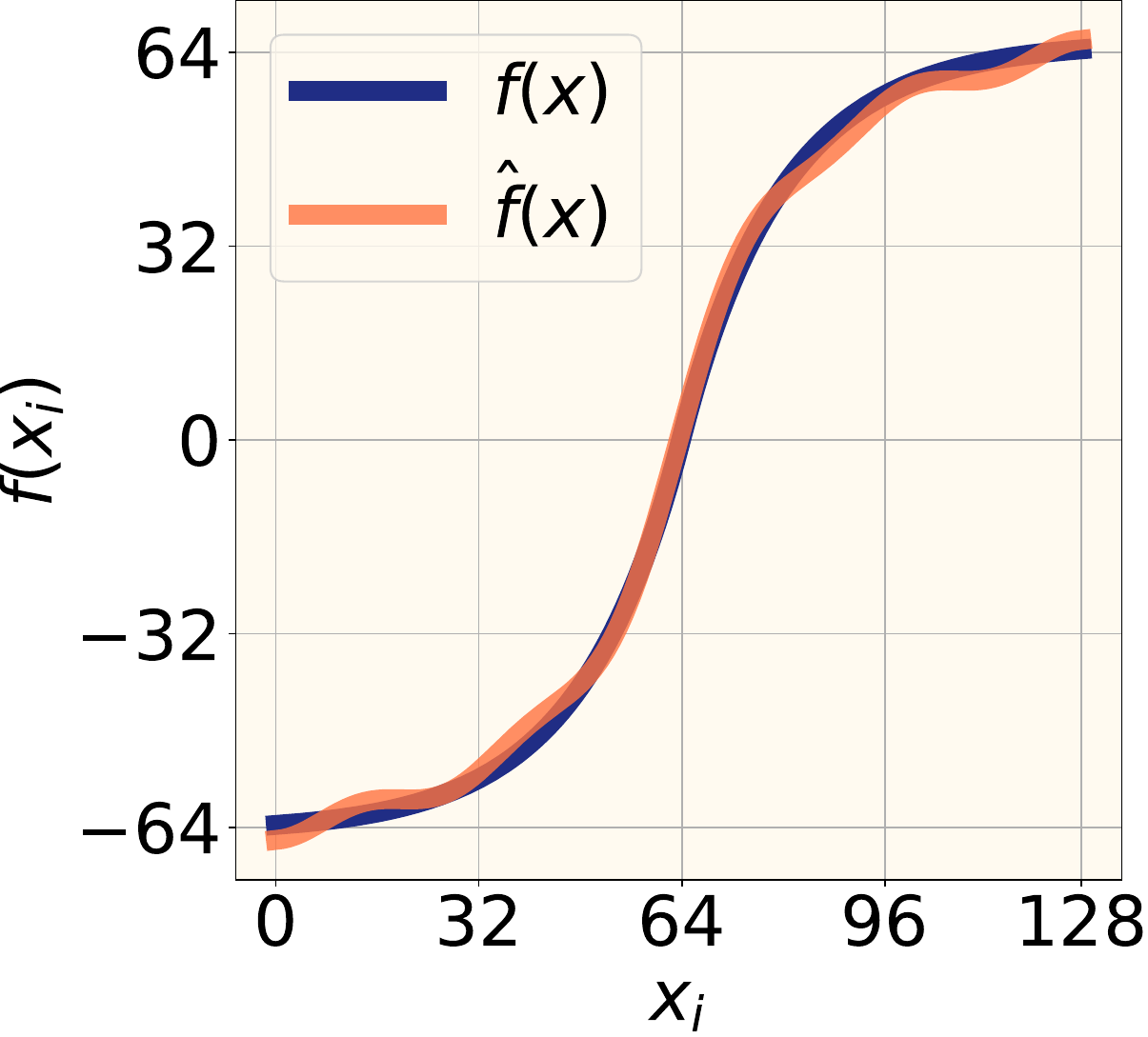}
    \end{subfigure}
    \hfill
    \begin{subfigure}[b]{0.305\textwidth}
        \includegraphics[width=\textwidth]{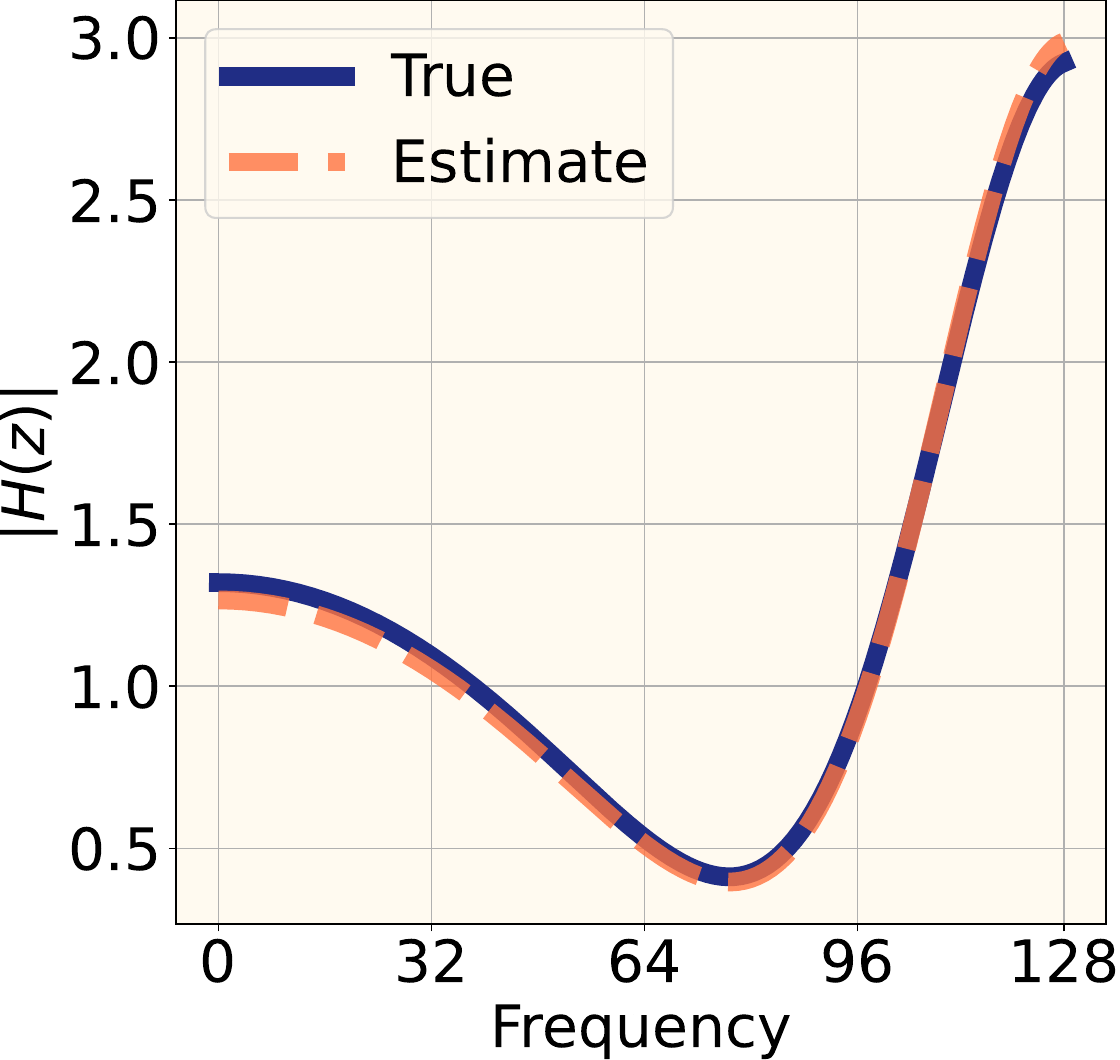}
    \end{subfigure}
    \hfill
    \begin{subfigure}[b]{0.32\textwidth}
        \includegraphics[width=\textwidth]{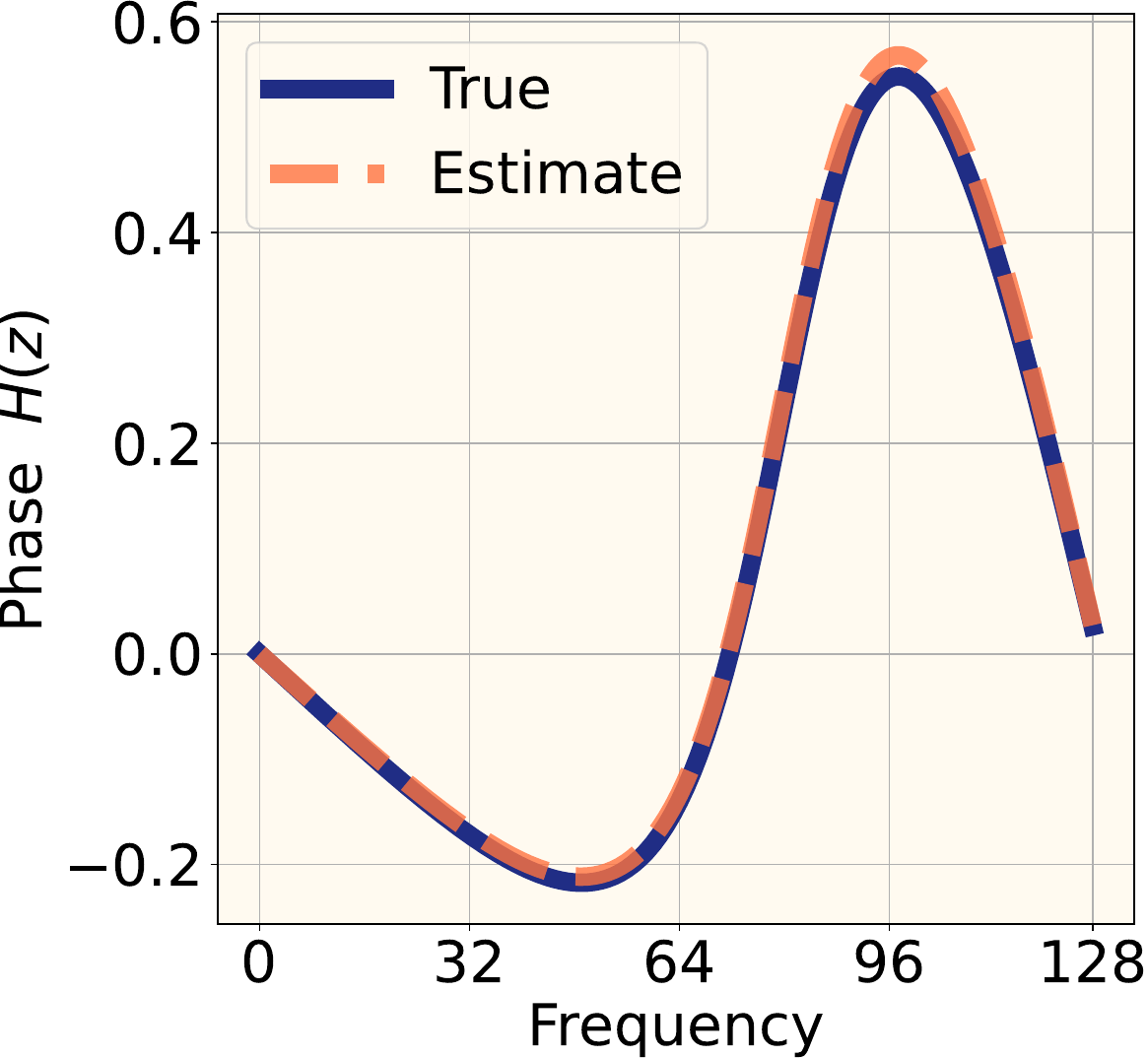}
    \end{subfigure}
    \subcaption{Compander nonlinearity and an IIR channel at $\text{SNR}^\text{pre}=\infty$.}
    \label{fig:iir_mdir_dct_snr_inf}
    \end{subfigure}
    \vspace{1em} 
    
    \begin{subfigure}[t]{\columnwidth}
    \begin{subfigure}[b]{0.315\textwidth}
        \includegraphics[width=\textwidth]{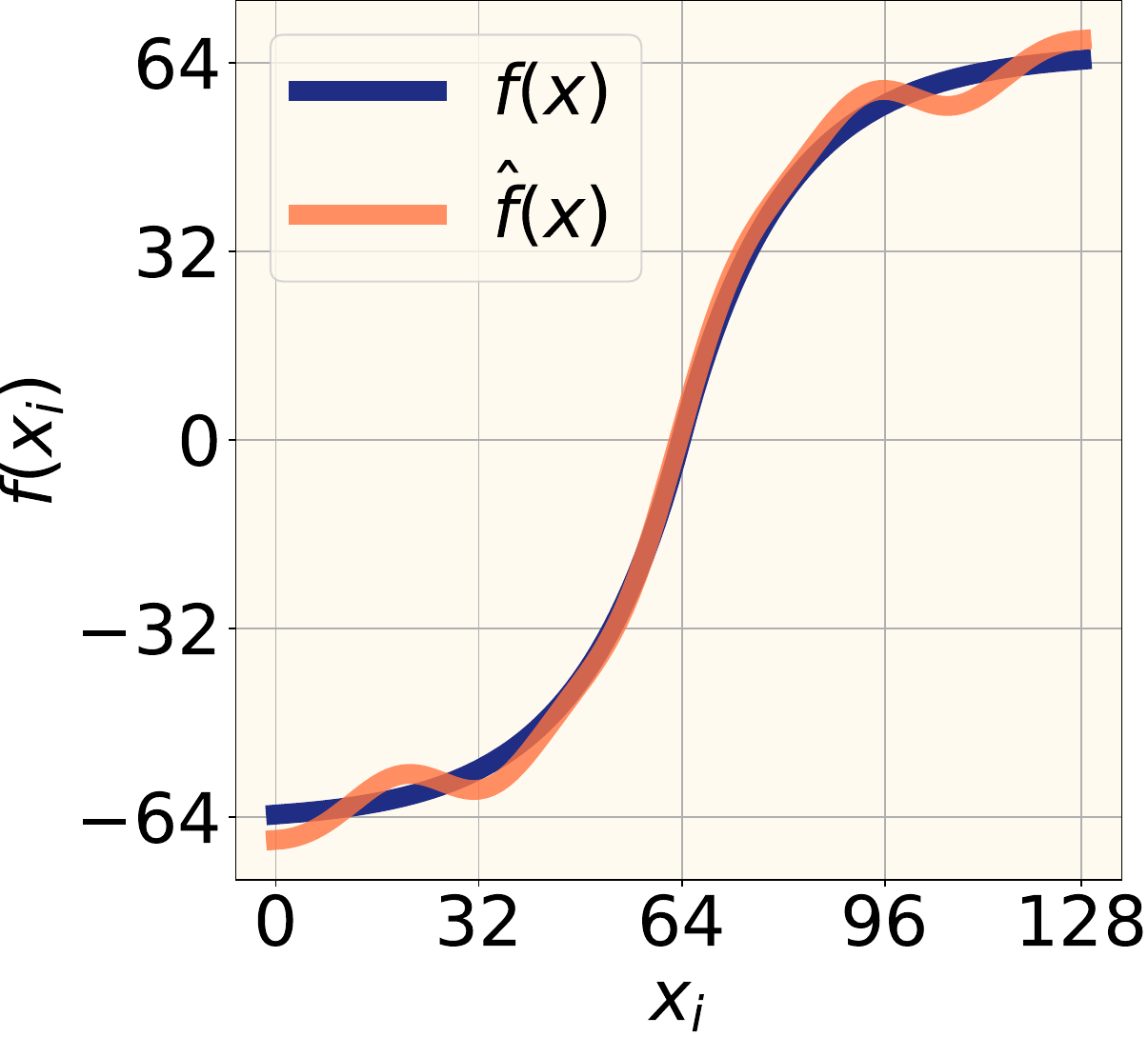}
    \end{subfigure}
    \hfill
    \begin{subfigure}[b]{0.305\textwidth}
        \includegraphics[width=\textwidth]{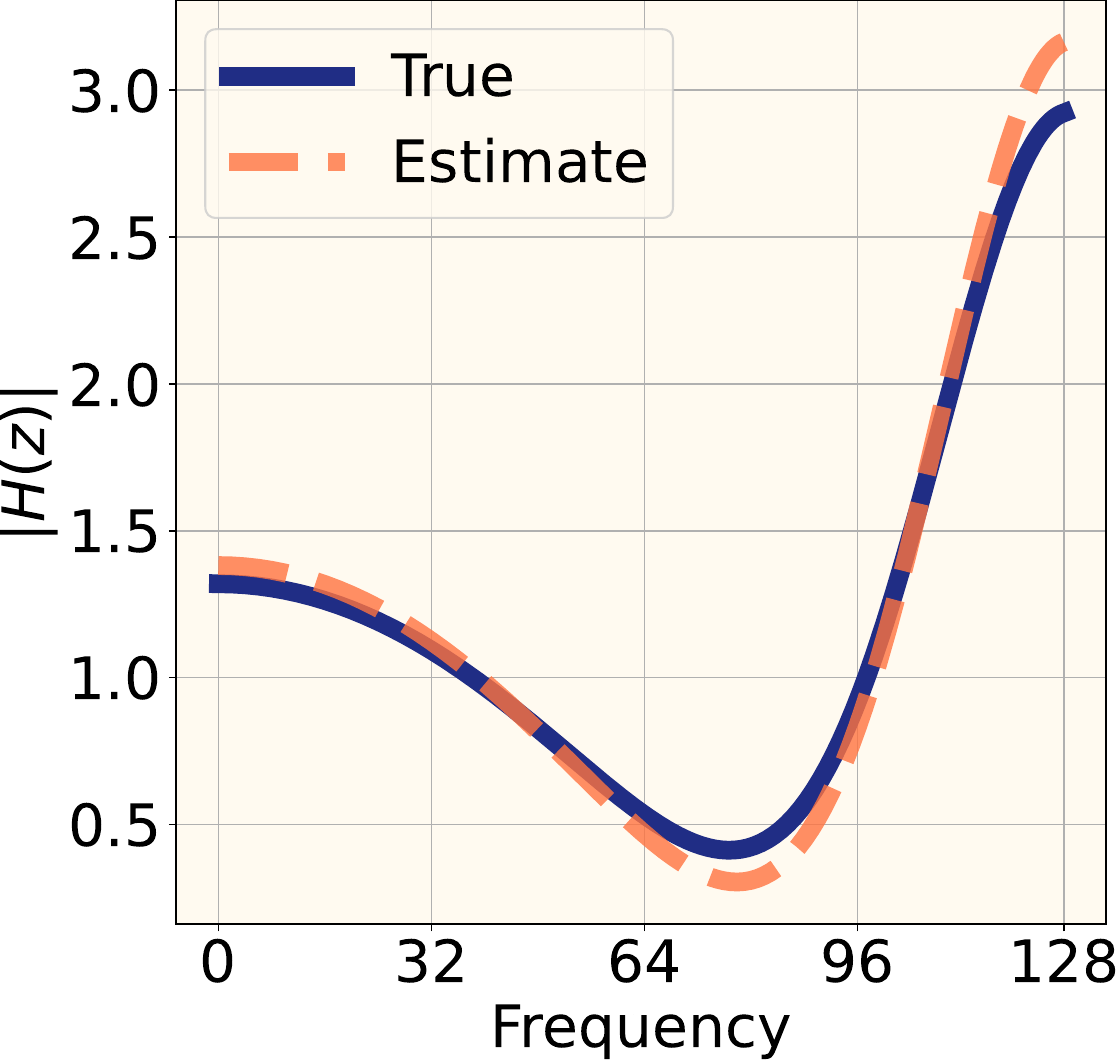}
    \end{subfigure}
    \hfill
    \begin{subfigure}[b]{0.32\textwidth}
        \includegraphics[width=\textwidth]{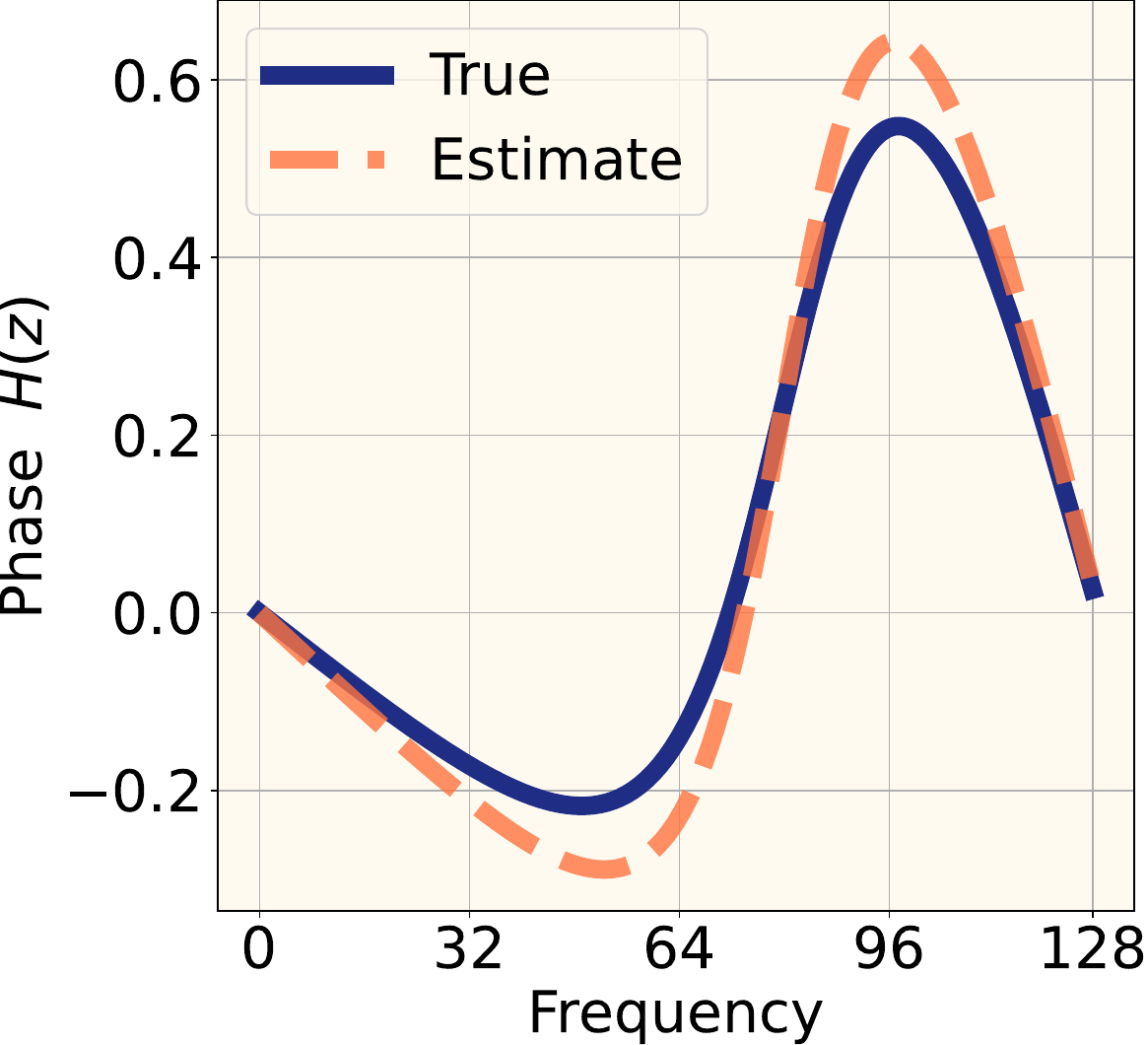}
    \end{subfigure}
    \subcaption{Compander nonlinearity and an IIR channel at $\text{SNR}^\text{pre}=10$ dB.}
    \label{fig:iir_mdir_dct_snr_10}
    \end{subfigure}
    \vspace{1em} 

    \begin{subfigure}[t]{\columnwidth}
    \begin{subfigure}[b]{0.315\textwidth}
        \includegraphics[width=\textwidth]{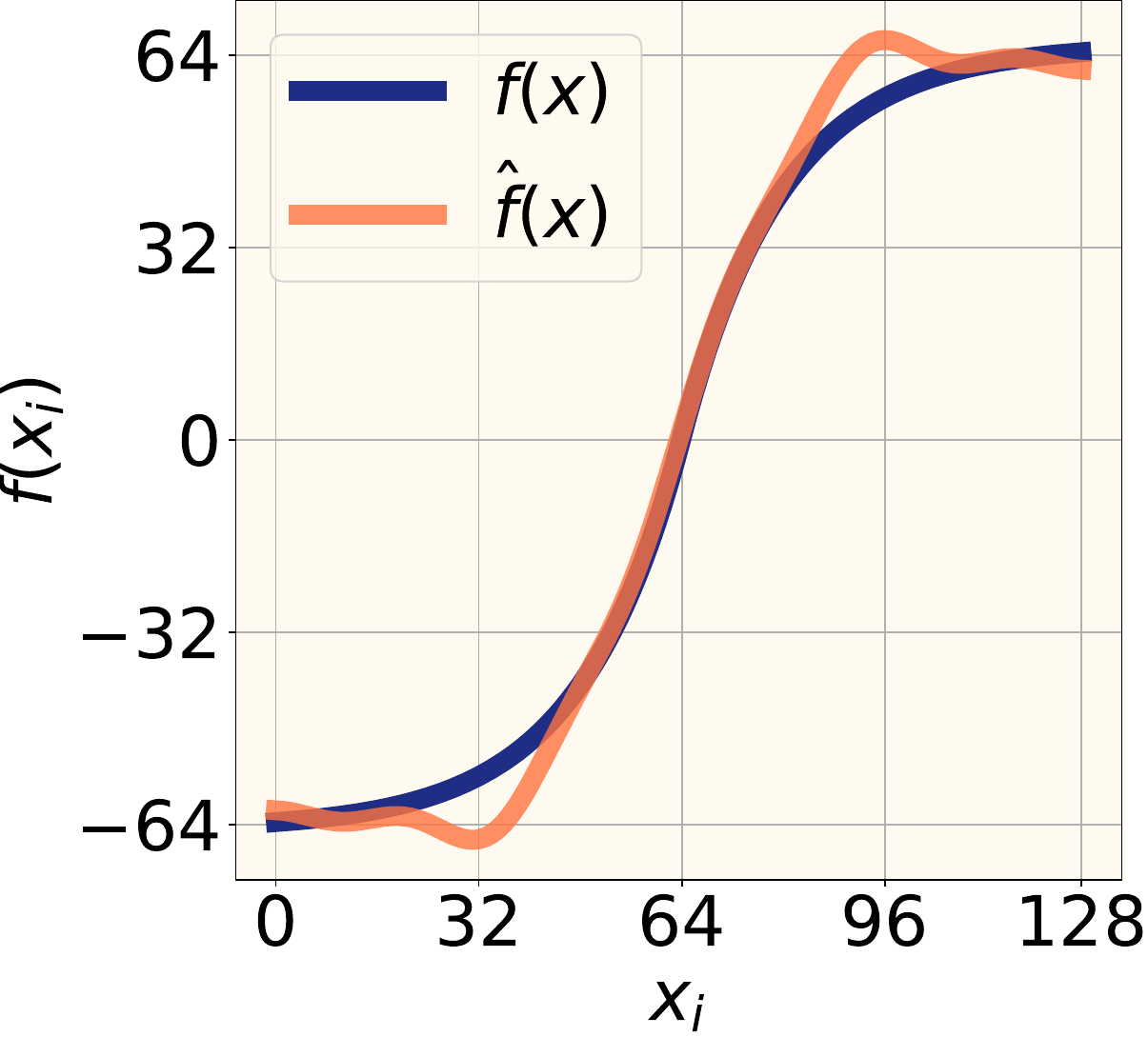}
    \end{subfigure}
    \hfill
    \begin{subfigure}[b]{0.287\textwidth}
        \includegraphics[width=\textwidth]{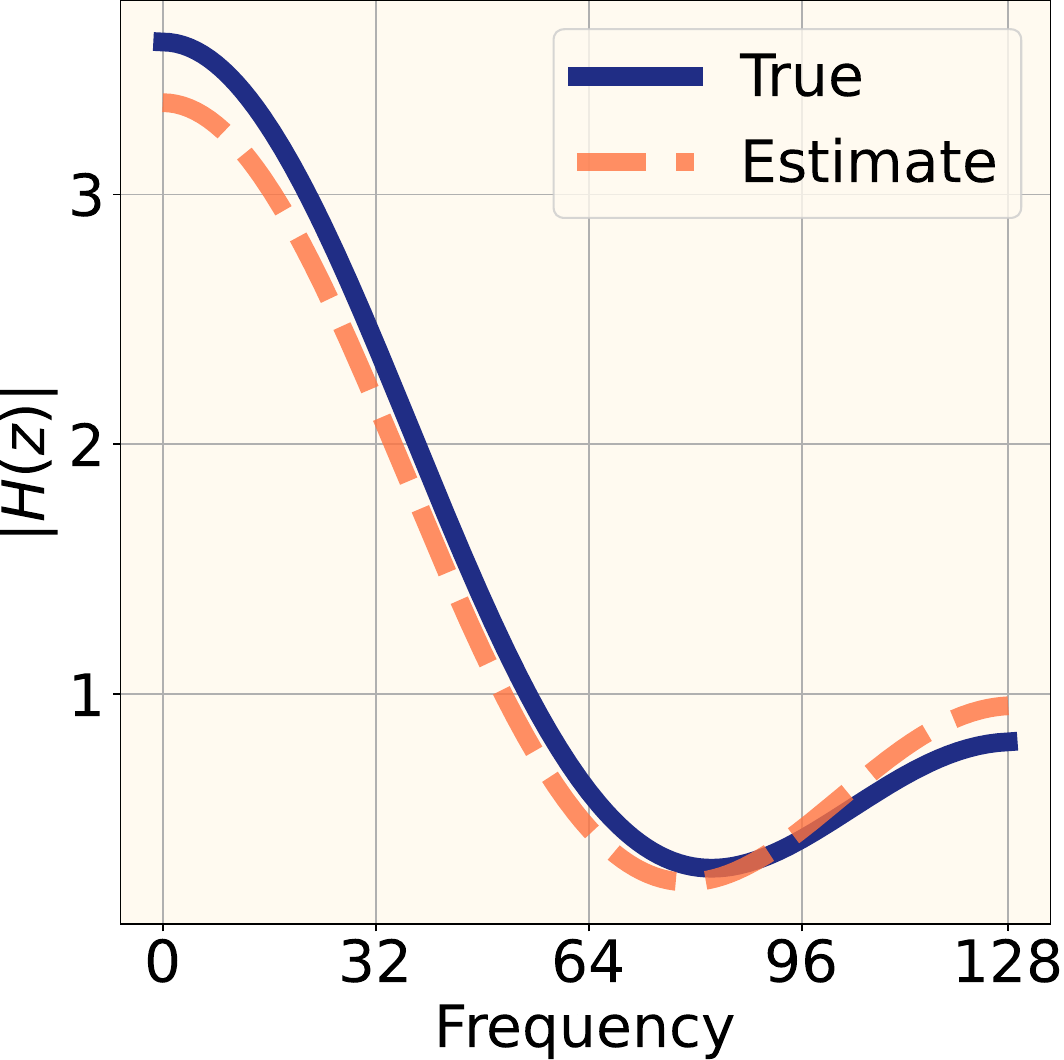}
    \end{subfigure}
    \hfill
    \begin{subfigure}[b]{0.32\textwidth}
        \includegraphics[width=\textwidth]{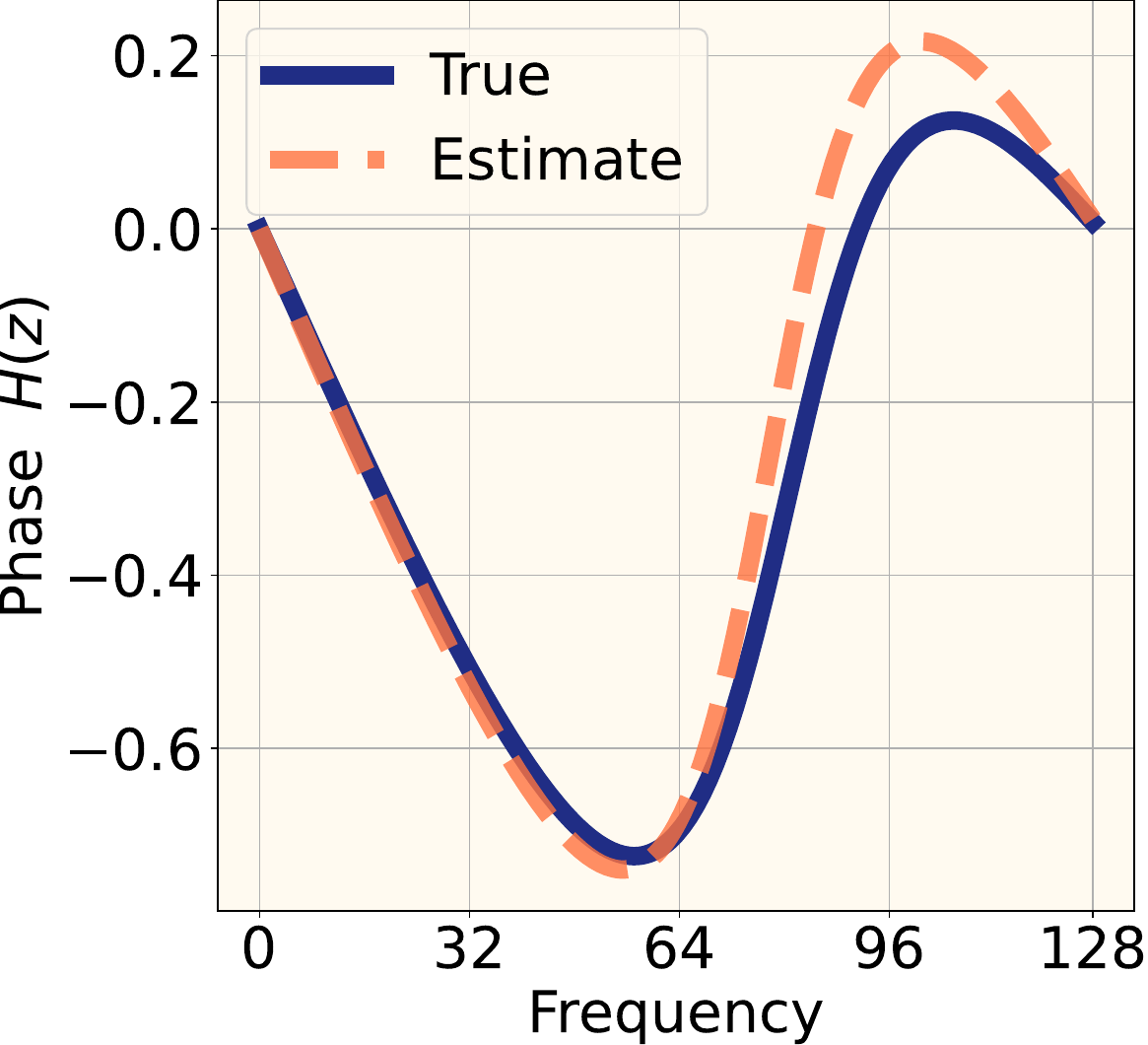}
    \end{subfigure}
    \subcaption{Compander nonlinearity and an FIR channel at $\text{SNR}^\text{pre}=10$ dB.}
    \label{fig:fir_mdir_dct_snr_10}
    \end{subfigure}

    \caption{Nonlinear frequency selective channel estimation for different configurations. We illustrate the nonlinear function, the channel magnitude and phase.}
    \label{fig:mdir_dct}
\end{figure}

Figure \ref{fig:mdir_dct}(\subref{fig:iir_mdir_dct_snr_inf}) shows the performance of the full estimation scheme in a noiseless setting. Each subfigure presents the true and estimated nonlinear function, along with the magnitude and phase response of the channel.
The proposed scheme accurately estimates all system parameters, confirming its effectiveness. This result supports the claim that, while the MDIR decomposition is inherently suboptimal, the MSE remains a valid and effective criterion for optimization even when combined with the nonlinear model. Furthermore, it provides empirical evidence that the alternating descent strategy converges reliably to a local minimum.

Figure \ref{fig:mdir_dct}(\subref{fig:iir_mdir_dct_snr_10}) illustrates the performance at 10 dB of SNR and Figure \ref{fig:mdir_dct}(\subref{fig:fir_mdir_dct_snr_10}) isolates the case where the channel is a \gls{FIR} ($\mathbf{b} = [1, 0, 0]^T$), which is a more realistic configuration of a radio communication channel. The proposed procedure performs reliably for SNR values above 10 dB, offering notable results in both the identification of the nonlinearity and the estimation of the channel.

\section{Conclusion}

This work has revisited one of the most fundamental problems in communications: nonlinear channel estimation, reconsidered from a modern perspective that reinterprets classical signal processing theory in light of current trends toward AI. The goal was not to simply propose another nonlinear estimator, but to initiate a broader reflection on how communication theory should evolve in an era increasingly dominated by AI, while preserving rigor and interpretability.

We have introduced a formal nonlinear signal processing framework rooted in the DCT, which provides a compact, interpretable and mathematically grounded representation of nonlinearities. Building on previous applications to nonlinear function approximation and fully-adaptive neural networks, this work extends the framework to nonlinear channel estimation. When embedded into a classical receiver architecture, the DCT enables data-driven learning of the channel response in a structured, controllable and transparent manner. This white-box approach reconciles adaptive learning with physical interpretability, demonstrating the versatility of the framework for nonlinear signal processing in communications. 
The results show that the MDIR-DCT combination can accurately model nonlinear channels, laying a solid foundation for future stages such as precompensation at the transmitter, receiver compensation, or advanced detection strategies.

The framework should be regarded as a starting point, highlighting that rigorous signal processing and learning are not opposing paradigms but complementary. On the contrary, they can, and should, coexist. 
Future research should explore richer nonlinear representations, advanced compensation architectures and detection strategies that extend this foundation to more complex channels and multiuser systems.

Overall, the central message is that progress in communications requires more than transferring AI paradigms from other fields. Before AI takes over, we must ensure that the discipline preserves its theoretical rigor, physical insight and interpretability. The path forward lies not in abandoning decades of knowledge in signal processing and information theory, but in integrating data-driven learning into these well-established frameworks. The proposed DCT-based approach represents a step in this direction, one that invites the community to revisit existing techniques and rethink nonlinear signal processing from the ground up, bridging the precision of theory with the adaptability of learning.


\section{Author Contributions}
AP has conceived, structured the paper, as well as written the main part and supervised its whole writing.
MM has written and organized the simulation part.
ML has conceived the DCT-based nonlinear signal processing.
The corresponding author is MM.

\bibliographystyle{IEEEtran}
\bibliography{refs}

\end{document}